\documentclass[5p,authoryear]{elsarticle}
\usepackage{amssymb}
\usepackage{amsmath}
\usepackage{url}
\usepackage{caption}
\usepackage{subcaption}
\usepackage{graphicx}
\usepackage{color}
\usepackage{microtype}
\usepackage{booktabs}
\usepackage{verbatim}

\journal{Astronomy and Computing}

\begin{document}

\newcommand{\code}[1]{\texttt{#1}}
\newcommand{\degrees}[1]{$#1^\circ$}

\newcommand{\githash}{69ae54a}\newcommand{\gitdate}{2015-03-02}

\begin{frontmatter}

\title{The LOFAR Transients Pipeline\tnoteref{git}}
\tnotetext[git]{Generated from git source \texttt{\githash} dated \gitdate.}

\author[pton,api]{John~D.~Swinbank}
\ead{swinbank@princeton.edu}
\author[oxford]{Tim~D.~Staley}
\author[api,rhodes]{Gijs~J.~Molenaar}
\author[api]{Evert~Rol}
\author[cass]{Antonia~Rowlinson}
\author[cwi,api]{Bart~Scheers}
\author[nki]{Hanno~Spreeuw}
\author[cass,arc]{Martin~E.~Bell}
\author[oxford,soton]{Jess~W.~Broderick}
\author[api]{Dario~Carbone}
\author[cea]{Hugh~Garsden}
\author[api]{Alexander~J.~van~der~Horst}
\author[ucb]{Casey~J.~Law}
\author[astron]{Michael~Wise}

\author[jodrell,soton]{Rene~P.~Breton}
\author[api]{Yvette~Cendes}
\author[cea,nancay]{St\'ephane~Corbel}
\author[thur]{Jochen~Eisl\"offel}
\author[radboud,astron]{Heino~Falcke}
\author[oxford]{Rob~Fender}
\author[lpc2e,nancay]{Jean-Mathias~Grie{\ss}meier}
\author[astron,api]{Jason~W.T.~Hessels}
\author[jodrell]{Benjamin~W.~Stappers}
\author[oxford]{Adam~J.~Stewart}
\author[api]{Ralph~A.M.J.~Wijers}
\author[api]{Rudy~Wijnands}
\author[lesia]{Philippe~Zarka}

\address[pton]{Department of Astrophysical Sciences, Princeton University, Princeton, NJ 08544, USA}
\address[api]{Anton Pannekoek Institute, University of Amsterdam, Postbus 94249, 1090 GE Amsterdam, The Netherlands}
\address[oxford]{Astrophysics, Department of Physics, University of Oxford, Keble Road, Oxford OX1 3RH, UK}
\address[rhodes]{Department of Physics and Electronics, Rhodes University, PO Box 94, Grahamstown, 6140 South Africa}
\address[cass]{CSIRO Astronomy and Space Science, PO Box 76, Epping, NSW 1710, Australia}
\address[cwi]{Centrum Wiskunde \& Informatica, PO Box 94079, 1090 GB Amsterdam, The Netherlands}
\address[nki]{Netherlands Cancer Institute, Postbus 90203, 1006 BE Amsterdam, The Netherlands}
\address[arc]{ARC Centre of Excellence for All-sky Astrophysics (CAASTRO), The University of Sydney, NSW 2006, Australia}
\address[soton]{Department of Physics and Astronomy, University of Southampton, Southampton SO17 1BJ, UK}
\address[cea]{Laboratoire AIM (CEA/IRFU - CNRS/INSU - Universit\'e Paris Diderot), CEA DSM/IRFU/SAp, F-91191 Gif-sur-Yvette, France}
\address[ucb]{Department of Astronomy and Radio Astronomy Lab, University of California, Berkeley, CA, USA}
\address[astron]{ASTRON, The Netherlands Institute for Radio Astronomy, Postbus 2, 7990 AA Dwingeloo, The Netherlands}
\address[jodrell]{Jodrell Bank Centre for Astrophysics, School of Physics and Astronomy, The University of Manchester, Manchester M13 9PL, UK}
\address[nancay]{Station de Radioastronomie de Nan\c{c}ay, Observatoire de Paris, CNRS/INSU, USR 704 - Univ. Orl\'eans, OSUC, 18330 Nan\c{c}ay, France}
\address[thur]{Th\"uringer Landessternwarte, Sternwarte 5, D-07778 Tautenburg, Germany}
\address[radboud]{Department of Astrophysics/IMAPP, Radboud University Nijmegen, PO Box 9010, 6500 GL Nijmegen, The Netherlands}
\address[lpc2e]{LPC2E - Universit\'{e} d'Orl\'{e}ans / CNRS, 45071 Orl\'{e}ans cedex 2, France}
\address[lesia]{LESIA, Observatoire de Paris, CNRS, UPMC, Universit\'e Paris-Diderot, 5 place Jules Janssen, 92195 Meudon, France}

\begin{abstract}

Current and future astronomical survey facilities provide a remarkably rich
opportunity for transient astronomy, combining unprecedented fields of view
with high sensitivity and the ability to access previously unexplored
wavelength regimes. This is particularly true of LOFAR, a
recently-commissioned, low-frequency radio interferometer, based in the
Netherlands and with stations across Europe. The identification of and
response to transients is one of LOFAR's key science goals.  However, the
large data volumes which LOFAR produces, combined with the scientific
requirement for rapid response, make automation essential. To support this, we
have developed the LOFAR Transients Pipeline, or TraP. The TraP ingests
multi-frequency image data from LOFAR or other instruments and searches it for
transients and variables, providing automatic alerts of significant detections
and populating a lightcurve database for further analysis by astronomers.
Here, we discuss the scientific goals of the TraP and how it has been designed
to meet them. We describe its implementation, including both the algorithms
adopted to maximize performance as well as the development methodology used to
ensure it is robust and reliable, particularly in the presence of artefacts
typical of radio astronomy imaging. Finally, we report on a series of tests of
the pipeline carried out using simulated LOFAR observations with a known
population of transients.

\end{abstract}


\end{frontmatter}

\section{Introduction}
\label{sec:intro}

\subsection{Slow transients and variable sources}
\label{sec:intro:science}

While most objects in the Universe are steady on human timescales, there are
classes of sources displaying variability on timescales of years to days,
seconds or fractions of a second. This variability can occur regularly, at
irregular intervals, or as a singular event from a given object. Searching
for these transients and variables requires observatories with a large field
of view, a capability which was up to now reserved only for some optical
telescopes and X- and $\gamma$-ray satellites. The radio regime is now also
entering this area of time-domain astronomy, with several new facilities being
built that have large fields of view (several square degrees or larger) and
transients as one of their key scientific objectives
\citep[e.g.,][]{Taylor:2012,Murphy:2013,Tingay:2013,vanHaarlem:2013,Bell:2014}.
A few of these observatories are probing the low radio frequency regime, from
tens to hundreds of MHz, a range that has been largely unexplored so far.

Transients in the low-frequency radio sky can be divided into roughly two
classes, characterized by their emission processes and the observing
techniques used to study them: coherent emitters, which display very fast
variability, and are found mostly in beamformed time series data, and
incoherent emitters, which display slow variability and are usually detected
by comparing multiple images of the same field \citep{Fender:2011}. Here, we
take the dividing line between slow and fast as $\sim1$~second, which is the
fastest time scale at which radio images are typically made \citep[but see
e.g.][]{Law:2011}. The most well known examples of coherent emitters are
pulsars and masers, but coherent emission processes are also predicted, albeit
not yet discovered, for other sources like gamma-ray bursts \citep{Usov:2000,
Sagiv:2002} and magnetar flares \citep{Lyubarsky:2014}. This paper, however,
focuses on searching for incoherent transients and variable sources in the
image plane, on timescales from seconds to years.

The main incoherent emission process at low radio frequencies is synchrotron
radiation, which arises when relativistic electrons are accelerated in strong
magnetic fields. It is produced where a large amount of energy is injected
into the ambient medium in jet sources and explosive events, such as X-ray
binaries, active galactic nuclei, tidal disruption events, gamma-ray bursts,
supernovae, magnetars, and flare stars \citep[e.g.,][]{Dent:1965,Gregory:1972,
Frail:1997,Frail:1999,Levan:2011}. While many of these sources show short
bursts of emission at X- or $\gamma$-ray energies, their variability timescale
at low radio frequencies is much longer, because the radiative lifetimes of
the particles to synchrotron emission are very long, and due to synchrotron
self-absorption effects \citep{vanderLaan:1966}. Although the latter decreases
the sources' brightness, making their detection more challenging, it has a
high pay-off scientifically since determining the evolution of the spectrum at
low radio frequencies provides important information on the energetics
involved in these events, the acceleration of electrons up to relativistic
velocities, the generation of magnetic fields, the production and collimation
of jets, and the feedback of these jets on their surroundings. Furthermore,
sources with small angular scales on the sky, like active galactic nuclei,
show variability which is not intrinsic but caused by scattering in the
interstellar medium \citep{Rickett:1990}. Therefore these sources are not
only interesting for studying their physical properties, but can also be used
to probe the medium in between them and us. In this context we note that some
coherent events that are intrinsically very short can be scattered and
dispersed in the interstellar medium, smearing out their signal to timescales
that are probed by image transient searches \citep[see
e.g.][]{Broderick:2014}.

\subsection{Detecting transients and variables}
\label{sec:intro:tech}

The transient sky has long been studied across the electromagnetic spectrum,
but the scale of transient searches has increased markedly recently, in
particular in the optical and radio regimes.

Searching for transients with large field-of-view X- and $\gamma$-ray
instruments has been common for a long time, and a variety of techniques have
been used for all-sky monitors on board the {\it Rossi X-ray Timing Explorer}
\citep{Levine:1996}, {\it Compton Gamma Ray Observatory} \citep{Fishman:1992},
{\it Swift} \citep{Gehrels:2004}, and {\it Fermi Gamma-ray Space Telescope}
\citep{Atwood:2009,Meegan:2009}. The most common way to find rapid transients at
these energies is by monitoring a large fraction of the sky, and triggering on
a sudden increase in the total X- or $\gamma$-ray flux. Alternative
techniques are required for transients that evolve more slowly: for instance,
the Earth occultation method described by \citet{Harmon:2002}.

Transient searches in the image domain over similarly large fields-of-view are
now planned---and, indeed, already being carried out---at optical and radio
frequencies. Here, efficiently searching the extremely large data volumes
produced is challenging. Optical telescopes optimized to search for transients
include the Catalina Real-Time Transient Survey \citep{Drake:2009}, Palomar
Transient Factory \citep{Rau:2009}, Pan-STARRS \citep{Denneau:2013}, and the
Las Cumbres Observatory Global Telescope Network \citep{Brown:2013}.  Several
radio telescopes have dedicated transient programs as well, notably the Jansky
Very Large Array (JVLA), AMI \citep{Staley:2013}, MeerKAT \citep[Karoo Array
Telescope;][]{Booth:2012}, ATA \citep[Allen Telescope Array;][]{Welch2009}
ASKAP \citep[the Australian Square Kilometre Array
Pathfinder;][]{Murphy:2013}, the MWA \citep[Murchison Widefield
Array;][]{Tingay:2013,Bell:2014}, the LWA \citep[Long Wavelength
Array;][]{Taylor:2012} and LOFAR \citep[the Low Frequency
Array;][]{vanHaarlem:2013}. In the longer term, the Large Synoptic Survey
Telescope \citep[LSST;][]{Ivezic:2014} in the optical and Square Kilometre
Array \citep[SKA;][]{Dewdney:2010} will produce a dramatic increase in the
number of transients which can be detected.

Broadly, there are two possible techniques which are adopted by these
searches: difference imaging \citep[e.g.][]{Alard:1998,Law:2009} or on a
comparison of a list of sources measured in a given image against a deep
reference catalogue \citep[e.g.][]{Drake:2009}.

Difference imaging has been demonstrated to be effective when applied to
optical data, particularly in crowded fields which would suffer from
source-confusion in a catalogue-based survey. However, the efficacy of
difference imaging in the optical is partly due to to the sources of noise
being relatively well characterised, with pixel-noise largely independently
distributed and occurring on a different spatial scale to real sources
(assuming a well-sampled point-spread function), and the fact that optical
survey point-spread functions usually vary in a smooth fashion amenable to
model-fitting.

In contrast, noise in radio-synthesis images is inherently correlated on
similar scales to the sources of interest. Furthermore, effects such as radio
frequency interference (RFI) and interaction between faint beam-sidelobes and
bright out-of-field sources may cause artefacts which are harder to
characterise and correct for than those found in optical data. As a result,
higher signal-to-noise thresholds are typically applied to ensure that most
spurious detections are rejected \citep[although this process remains
fallible;][]{Frail:2012}. This degrades the sensitivity advantage of
the difference imaging technique, and so a cataloguing survey provides
equivalent results with the added benefit of recording lightcurves for
individual sources.

Many recent developments, including the precursors of this work, focus on the
latter approach: compiling lightcurves, storing them in a database, and then
searching for transients with a variety of statistical techniques
\citep{Spreeuw:2010,Bannister:2011,Bower:2011,Croft:2011,Swinbank:2011,
Thyagarajan:2011,Banyer:2012,Hancock:2012,Williams:2013,Mooley:2013,Bell:2014}.
The same strategy has been adopted in this work, which describes the pipeline
and methods developed for searching for transients with LOFAR\@. The system
described here also has a broader applicability to other instruments and is
developed with an eye to the long-term requirements of the SKA\@.

\section{Transients with LOFAR}
\label{sec:lofar}

\subsection{The International LOFAR Telescope}
\label{sec:lofar:ilt}

LOFAR is a recently-commissioned radio interferometer based in the
Netherlands and with stations across
Europe\footnote{\url{http://www.astron.nl/~heald/lofarStatusMap.html}}. LOFAR
operates in the low-frequency radio regime, observing in the frequency ranges
10--80 (low band) and 110--240 (high band) MHz, corresponding to wavelengths
between 30 and 1.2 metres. The system pioneers the concept of a
\textit{software telescope}, as signals received by simple crossed-dipole
antennas, which are sensitive to the whole sky, are digitized as soon as
possible and signal processing is done in software. In the low band the
voltages from each dipole are directly digitized; in the high band, an
analogue combination of data from 16 antenna elements (a ``tile'') is formed
prior to digitization.

In a typical imaging observation, each station acts as a phased array: the
digitized data from each of the antennas in the station is coherently combined
(``beamformed'') to point the station in a particular direction. The field of
view of beams formed in this way depends on frequency and station
configuration, with the full width at half maximum ranging from
19.55$^{\circ}$ for a core Dutch LOFAR station at 30\,MHz to 1.29$^{\circ}$
for an international LOFAR station at 240\,MHz\footnote{The various types of
LOFAR stations together with their key parameters are listed at
\url{http://www.astron.nl/radio-observatory/astronomers/lofar-imaging-capabilities-sensitivity/lofar-imaging-capabilities/lofa};
see \citet{vanHaarlem:2013} for detailed information on LOFAR's
configuration.}.  This beamformed data is then transported to the central
processing (CEP) facility at the University of Groningen where it is
correlated by a GPU-based system.

The software-based processing model provides for a great deal of flexibility.
After digitization, a polyphase filter splits the data into 0.2\,MHz wide
subbands, the number of subbands depending on the quantization of the data: it
is possible to trade off dynamic range for increased subband number and hence
bandwidth: in ``16 bit mode'', 244 subbands are available; in ``8 bit mode'',
488. These subbands can be spread across frequency space, to give a large
observing bandwidth (48.8\,MHz in 16 bit mode).  Alternatively, the beamformer
can be configured to form multiple beams with different selections of subbands
in different directions. In this latter mode, by trading off against observing
bandwidth, an extremely wide field of view may be observed.

When operating in imaging mode, the correlator provides a dump time of 1\,s,
and it is this which provides a lower-limit to the timescales which can be
searched for image plane transients. An alternative is to directly search high
time resolution beamformed data for fast transients, as described by
\citet{Stappers:2011} and \citet{Coenen:2014}. It is also possible to
configure the telescope such that beamformed and image data is recorded
simultaneously, providing the greatest possible range of cadences in a
transient search; ultimately, continual monitoring in this mode is an
important goal.

\subsection{The Transients Key Science Project and the Radio Sky Monitor}
\label{sec:lofar:rsm}

LOFAR's development and commissioning have been driven by six science areas:
the epoch of reionization, deep extragalactic surveys, cosmic magnetism, solar
physics and space weather, cosmic rays, and transient and variable sources.
The last of these is the remit of the Transients Key Science
Project\footnote{\url{http://www.transientskp.org/}}
\citep[TKSP;][]{Fender:2006}. The TKSP's interests include transient and
variable sources on all timescales, from sub-second changes in beamformed data
\citep{Stappers:2011} to multi-year variability monitored through long-term
imaging programmes; see \citet{vanHaarlem:2013} and references therein for a
complete discussion of the TKSP science case. It is upon detection and
monitoring in the image plane which this work concentrates.

\begin{figure}
  \begin{center}
    \includegraphics[width=0.8\columnwidth]{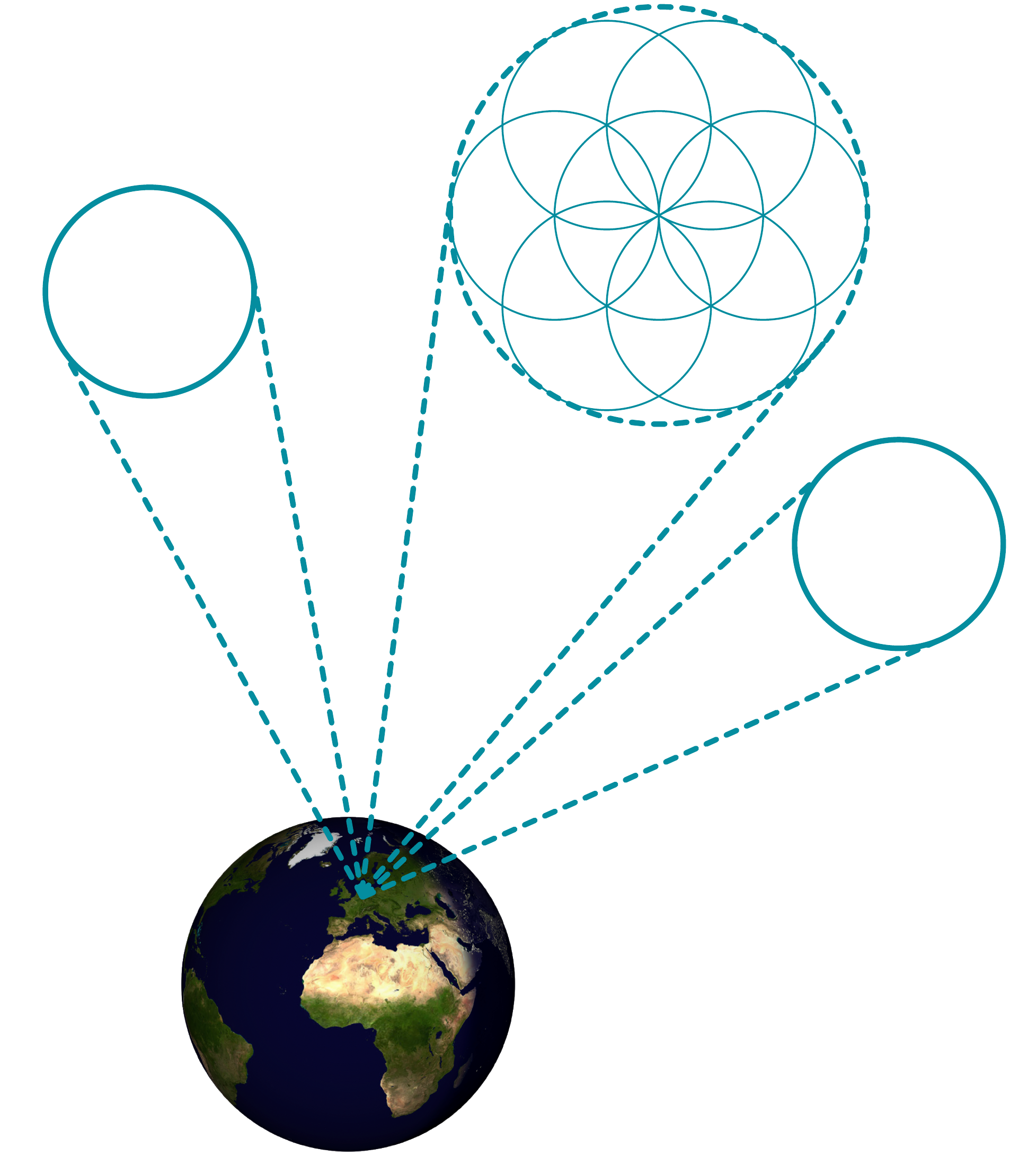}
  \end{center}
  \caption{The Radio Sky Monitor concept. Multiple LOFAR station beams tile
  out a large field of view, while other beams are available for target
  observations.}
  \label{fig:rsm}
\end{figure}

%
%

A key programme for the TKSP is the ``Radio Sky Monitor'', or RSM
\citep{Fender:2014}. In this mode, multiple beams from LOFAR are used to tile
out a large area on the sky (Fig.~\ref{fig:rsm}). This field of view is then
imaged on a  logarithmic range of timescales, from 1 to 10000\,s, and at a
range of frequencies, and that image stream is searched for transient and
variable sources. The survey strategy is flexible, but most plausible
strategies will focus on the galactic plane and the zenith, while taking
advantage of the large field of view to ensure that a large fraction of the
sky is regularly monitored. While this procedure is ongoing, individual beams
can be diverted from the ongoing survey to monitor specific targets of
interest or respond to transient alerts in real time (although the latter is
currently not implemented).

There are two key data products which result from this RSM: an archive of the
lightcurves observed for all point sources in the LOFAR sky, and low-latency
alerts of transient events. It is the TKSP's policy that, in general, these
products will both be made available to the community at large.

While the RSM is running, a large volume of correlated (visibility) and image
data will be generated. It is regarded as impractical to archive all of this
data. Instead, an averaged version of the visibility data may be stored with
reduced time and/or frequency resolution, and thumbnail images of significant
detections recorded.

Ultimately, LOFAR is designed to provide low-latency ``streaming'' image data.
When this is available, the transient search may be run, and alerts produced,
in real time. At time of writing, however, this capability is still in
development. Instead, visibility data is stored to disk for later ``offline''
imaging. This non-real-time version of the system has been deployed for
LOFAR's initial operational phase. In this mode, visibility data collected by
LOFAR undergoes some initial post-processing at CEP before being delivered to
the TKSP\@. Project members then image this data on local hardware, before
running the images through a version of the transients detection system which
is optimized for offline use.  In this way, TKSP members are able to develop
our understanding of LOFAR's imaging capabilities and to test and commission
the transients detection and monitoring pipeline (or ``TraP'') in advance of
its full deployment as part of a real-time LOFAR system.

It is on this TraP system which this manuscript focuses. In
\S\ref{sec:overview} we provide an overview of its inputs, outputs and overall
design. In \S\ref{sec:components} we describe in detail the algorithms
employed by the key pipeline components, and in \S\ref{sec:products} describe
the data products the pipeline delivers. Section~\ref{sec:implement} describes
the pipeline implementation on a technical level. Section~\ref{sec:dev}
discusses the development approaches taken. In \S\ref{sec:test} we describe
testing carried out on the TraP with simulated datasets. Finally, in
\S\ref{sec:future}, we describe enhancements which are planned for future
releases.

The TraP was developed with the aim of finding and monitoring transients in
RSM-like data. However, it is worth emphasizing that it should be ultimately
applicable to a much wider range of instrumentation. For example, it is
planned to use the TraP to scan as much LOFAR imaging data as possible, in a
so-called ``piggyback'' mode. An early version of the TraP has already been
used in a study of archival VLA data \citep{Bell:2011}, while a variant will
also be deployed as the transient detection system for AARTFAAC \citep[the
Amsterdam-ASTRON Radio Transients Facility and Analysis Centre, an
all-visible-sky monitor operating commensally with LOFAR;][]{Prasad:2012}.
Other developments target the Arcminute Microkelvin Imager Large Array
\citep[AMI-LA;][]{Staley:2013, Anderson:2014}, a 15\,GHz aperture synthesis
aperture synthesis radio telescope near Cambridge in the UK, and
KAT-7/MeerKAT, SKA-precursor telescopes in the Karoo Desert, South Africa.
Further variants targeting optical data are also under consideration.

The TraP is available as open source software; for more details, refer to
\S\ref{sec:future} and the code
repository\footnote{\url{https://github.com/transientskp/tkp/}}.

\section{Pipeline overview}
\label{sec:overview}

The design goal of the TraP is to automatically and rapidly identify transient
and variable sources within a time-series of image data. These sources may
be identified in two ways:

\begin{itemize}

  \item{\textit{New detections} are sources which appear at a location where,
  in previous epochs, no source was seen;}

  \item{\textit{Variables} are sources which have been observed for multiple
  epochs and show significant variability in their lightcurves.}

\end{itemize}

Such sources are identified automatically by the TraP, based on no prior
knowledge. It is also possible for the user to specify the location of known
sources for monitoring. Variability metrics are retained for all sources, so
that decisions on what constitutes an `interesting source' may be made after
data-processing (\S\ref{sec:products}).

Since the TraP is ultimately designed to perform near real-time processing of
an image stream, we assume that after an image has been processed it is no
longer available for further analysis (modulo the system described in
\S\ref{sec:implement:mongodb}). Therefore, the TraP currently provides no
capability to look back at previously processed images in the light of new
data: it does not, for example, attempt to go back and check earlier images of
the same position after observing a new transient. Although retaining the full
image stream is unlikely to be practical for projects which generate
substantial data volumes, future versions of the TraP may include the
capability to generate and store an average of the input data, using this to
increase the depth of the survey and improve source characterization.

\subsection{Inputs}
\label{sec:overview:inputs}

The fundamental input to the trap is a time-series of three-dimensional (two
spatial, one frequency) image ``cubes''. These are generally assumed to be
produced by the LOFAR imaging pipeline \citep{Heald:2010, Heald:2011,
Heald:2014, vanHaarlem:2013}, however, as described in
\S\ref{sec:components:accessors}, the code is designed to be agnostic as to
the format and origin of the data being ingested.

In addition, the TraP may optionally be given a user-defined list of
monitoring positions. Measurements are made and stored for each such position
in each plane of every image cube ingested, regardless of whether the
automatic source-finding routines regard it as significant.

\subsection{Products}
\label{sec:overview:products}

The TraP is designed to produce two key data products:

\begin{itemize}

  \item{Near real-time alerts to the community and/or targeted at specific
  partners describing ongoing transient celestial events;}

  \item{An archival database of lightcurves for all astronomical point sources
  detected during pipeline processing together with information about their
  variability.}

\end{itemize}

The pipeline system is flexible enough to provide alerts in a variety of
formats, and it is therefore able to interoperate with whatever mechanisms
other facilities have in place for receiving notifications. For example, one
can imagine e-mail or SMS being convenient. However, development has focused
on the VOEvent system \citep{Seaman:2011} and its association distribution
networks \citep{Williams:2012}. These provide a flexible and convenient method
for widespread alert dissemination, which is described in detail in
\S\ref{sec:implement:voevent}.

In addition to these fundamental data products, the TraP may optionally store
a copy of all the image pixel data processed for future reference. This is not
required for the analysis performed by the TraP, but we have found it
convenient to maintain an archive of some or all of the images processed for
display purposes (e.g.\ using the interface described in
\S\ref{sec:products:db:banana}).

\subsection{Methods}

\begin{figure}
  \includegraphics[width=\columnwidth]{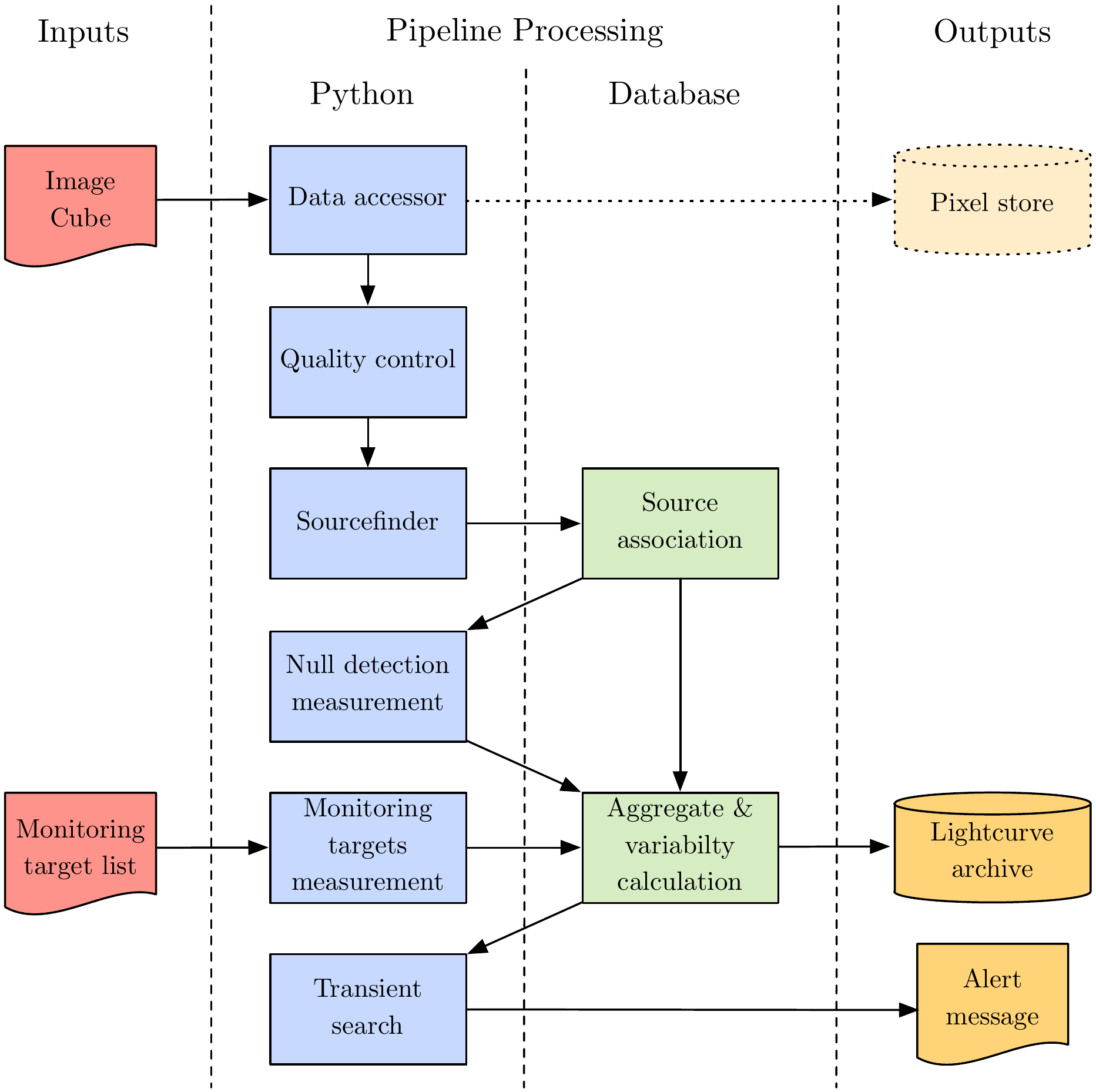}

  \caption{Transients Pipeline (``TraP'') overview, showing the flow of data
  through the system as an image ``cube'' is processed, ultimately producing
  an archival lightcurve database and (if appropriate) transient alert
  messages. The individual pipeline stages are described in
  \S\ref{sec:overview} and \S\ref{sec:components}; their implementation, in
  terms of a mixture of Python code and database routines, is discussed in
  \S\ref{sec:implement}. The dotted parts of the diagram represent optional
  functionality: they are not required for the core TraP functionality.}

  \label{fig:trap}
\end{figure}

To map from the inputs to the products described above, the following
procedure is adopted. Each of these stages is described in more detail in
\S\ref{sec:components}; their relationship is shown graphically in
Fig.~\ref{fig:trap}.

\begin{enumerate}

  \item{The image cube is stored in main memory as a series of two-dimensional
  images, each representing a particular frequency. The in-memory
  representation of an image used by the TraP is independent of the on-disk
  data storage format; see \S\ref{sec:components:accessors} for details.}

  \item{Each image undergoes a quality-control procedure, designed to identify
  and exclude from further processing data of unacceptably poor quality. Note
  that even when data is excluded by quality control it is not completely
  disregarded, but rather its existence and the reason for its rejection are
  recorded. For details of the quality control checks and the way in which
  they are applied see \S\ref{sec:components:qc}.}

  \item{A source-finding and measurement process is applied independently to
  each plane of the image cube. Sources which meet the relevant significance
  criteria are parameterized by elliptical Gaussians. For more details on the
  source finding procedure see \S\ref{sec:components:sf}.}

  \item{An ``association'' procedure is carried out, in which the new
  measurements are either identified as updates to the lightcurves of
  previously known sources, or as new, previously-undetected, sources. Details
  on the algorithms used for source association may be found in
  \S\ref{sec:components:assoc}.}

  \item{A list of sources which are expected to appear within the image but
  were \textit{not} detected by the source finding procedure above is now
  constructed following the procedure described in
  \S\ref{sec:components:nulldet}. The same source measurement code is now used
  to fit and record source parameters at each of these fixed positions, and
  the relevant lightcurves updated accordingly.}

  \item{The same source measurement code is used to fit and record source
  parameters at each of the user-specified monitoring positions, and the
  relevant lightcurves updated accordingly. This procedure is described in
  \S\ref{sec:components:monitoring}.}

  \item{For each lightcurve in the database a series of aggregate properties
  describing the astronomical source are calculated. These include a weighted
  mean position, flux density and a series of \textit{variability indices}
  which quantify the magnitude and significance of the variability of the
  lightcurve. This is described in \S\ref{sec:components:aggregate}.}

\end{enumerate}

At this point, the core loop of the pipeline is complete: the next (by time)
image cube in the sequence may be ingested and the process repeats. At the
same time, the results are immediately made available via a database, as
described in \S\ref{sec:products}.

Further analysis may be performed by querying the database for scientifically
relevant phenomena and reacting appropriately.  For example, one could search
for all bright sources which do not have a previously-detected counterpart,
and thereby identify new transients.  Alternatively, a query could search for
lightcurves which have reached a particular threshold in the variability
indices, or which meet some other user-defined criteria.

It is important to emphasize that these queries can be performed at any time.
For example, the user could wait until the complete pipeline run has been
completed and archived before searching the database; equally, however, a
real-time analysis system can query the database continuously as new results
are added, and thereby identify new transients immediately.

As new measurements are appended to the database, continuously-updated
measures such as the variability indices for a given lightcurve or the
weighted mean position of the associated astronomical source will change with
time. It is possible, therefore, that a particular source which was identified
as variable by the real-time analysis system at some particular timestep will,
in the fullness of time, be shown to not, in fact, vary significantly. In order
to ensure reproducibility, the database records all the intermediate values as
well as the final, archival result. That is, the user may query the archival
database not just for the eventual state of a particular source, but for its
state as recorded after the insertion of any particular image cube.

Finally, it should be noted that although it is possible to create LOFAR
images with full polarization, and notwithstanding the ultimate TraP design
goals, the current version of the TraP searches for transient and variable
sources only within total intensity (Stokes I) images, and other polarization
information is not used. In time, though, polarization information will be
essential for properly characterizing the sources being identified: see
\S\ref{sec:future} for more information on future plans.

\section{Key pipeline stages}
\label{sec:components}

In this section we describe the logical structure of the TraP, focusing on the
core stages of the pipeline and the algorithms that they employ.
Section~\ref{sec:implement} describes how this logical structure is
implemented in terms of deployed software and hardware resources.

\subsection{Data accessors}
\label{sec:components:accessors}

While the TraP has been developed with LOFAR in mind, many of the core issues
we are addressing are widely applicable to much of the emerging field of
transient astronomy. As such, we aim to make it easy to adapt the TraP to
ingest images from data sources other than LOFAR\@. The pipeline is designed
to be data-source agnostic: the origin of the data is abstracted away from the
scientific logic. This has proven to be useful as internal LOFAR data-storage
standards have evolved.

Data source abstraction is achieved by defining a uniform interface which all
routines in the pipeline use to access and manipulate image data. Derived
classes represent data from specific sources, providing different routines for
loading the data, translating telescope-specific metadata, and so on. Adding
support for data from a new telescope is generally straightforward: for most
instruments, just a few simple extensions to the predefined routines for
working with FITS\footnote{\url{http://fits.gsfc.nasa.gov/}} or
CASA\footnote{\url{http://casa.nrao.edu/}}/\texttt{casacore}\footnote{\url{http://casacore.googlecode.com/}}
are required.

This system has enabled the TraP to be used in conjunction with data not only
from LOFAR but also from the VLA and AMI-LA, as described in
\S\ref{sec:lofar:rsm}.

\subsection{Quality control}
\label{sec:components:qc}

During the imaging procedure and the future real-time imaging pipeline,
extremely large numbers of images will be produced for processing by the TraP.
Some of these images will be of insufficient standard for transient searches:
for instance, high RFI or poor calibration solutions can lead to increased
noise in the image or artefacts that may be mistaken as transients.

The quality control procedure identifies and rejects those images which do
not pass a range of tests. The system is modular: new tests can easily be
added as required. Further, tests may be made specific to certain
instrumentation by building upon the data accessor framework
(\S\ref{sec:components:accessors}).

The standard checks supplied with the released version of the TraP are all
LOFAR specific. They are:

\begin{itemize}
  \item{Test that the measured noise in the image does not significantly
  exceed the theoretically expected value (\S\ref{sec:components:qc:noise});}

  \item{Test for appropriate sampling and shape of the restoring beam
  parameters (\S\ref{sec:components:qc:beam});}

  \item{Test for proximity of the image pointing direction to bright radio
  sources (\S\ref{sec:components:qc:proximity}).}
\end{itemize}

An image which fails one or more of these tests is not further processed.
Details of the failure are logged for future reference.

These tests are designed to provide a quick and simple mitigation of common
failures observed during development and commissioning. As TraP moves into
production deployments, it will be possible to supplement them with a range of
more elaborate tests as and when required.

\subsubsection{Check for noisy images}
\label{sec:components:qc:noise}

A clear signature of a poor quality image is when the measured noise level
significantly differs from the theoretically expected value: measured values
which are either too low or too high are indicative of problems with the
observation or its calibration.

The theoretical noise in LOFAR images can be calculated using parameters
extracted from the image metadata, such as the array configuration and
integration time used \citep{Nijboer:2009, vanHaarlem:2013}.

To measure the observed RMS in an image, we conduct the following steps:

\begin{enumerate}

  \item{Select a square region congruent with the image centre and comprising
  25\,\% of the total pixels in the image;}

  \item{Iteratively reject pixel values more than $n$ standard deviations from
  the median, where $n$ is some user-defined parameter (typically four), until
  no further pixels are being rejected.}

  \item{Using the remaining pixels, calculate the mean pixel value and the
  RMS scatter around this mean value.}

\end{enumerate}

We then calculate a simple ratio between the measured RMS noise and the
theoretical noise. The image is rejected when this ratio falls outside a
user-specified range.

\subsubsection{Check restoring beam parameters}
\label{sec:components:qc:beam}

\begin{figure}
  \begin{subfigure}[b]{0.49\columnwidth}
    \includegraphics[width=\textwidth]{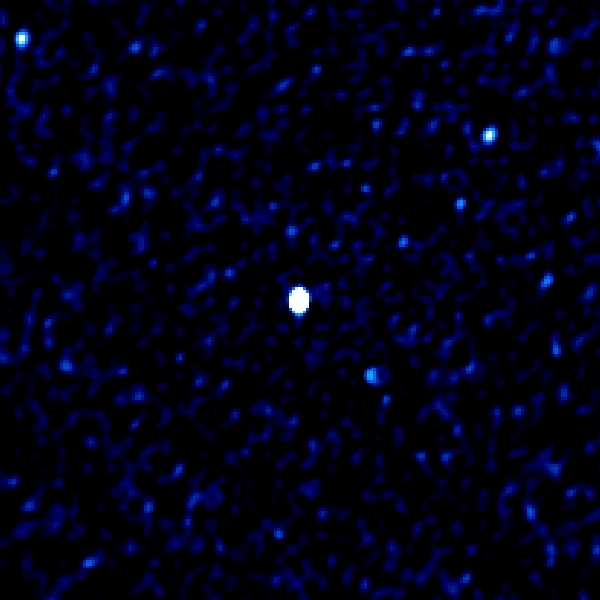}
    \caption{Correct sampling.}
    \label{fig:restoring_beam:normal}
  \end{subfigure}
  \hfill
  \begin{subfigure}[b]{0.49\columnwidth}
    \includegraphics[width=\textwidth]{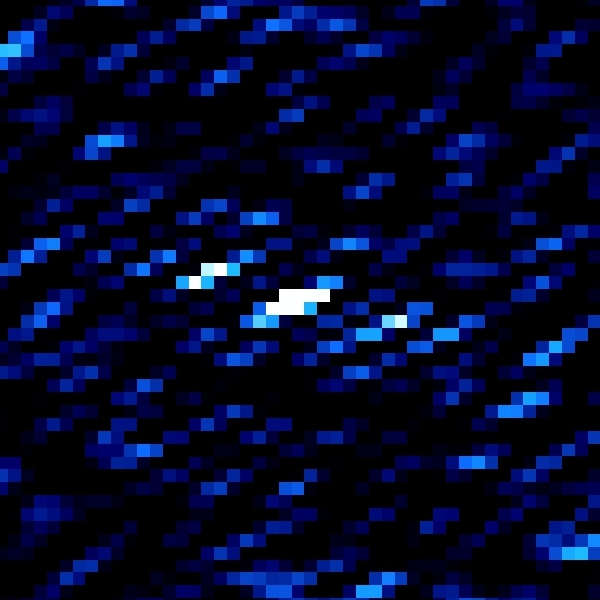}
    \caption{Inappropriate sampling.}
    \label{fig:restoring_beam:under}
  \end{subfigure}

  \caption{The effects of inappropriately sampling the restoring beam on image
  quality. These images are of the same field, centred on 3C\,295, and
  they share the same physical and colour scales. At left, the image is
  correctly sampled, with 3 pixels across the minor axis of the restoring
  beam.  To the right, only 1.3 pixels have been used: this image is
  unsuitable for pipeline processing.}

\label{fig:restoring_beam}
\end{figure}

The properties of the restoring beam \citep{Hogbom:1974} used to create the
images used within the TraP also play a significant role in assessing the
image quality. The image should be created such that the beam is appropriately
sampled, with around three pixels across its minor axis. Incorrect sampling
can cause increased noise, artefacts and spurious source detections, as
illustrated in Fig.~\ref{fig:restoring_beam}. The TraP aims to be robust
against this, regardless of the origin of the images.

The shape of the beam is also considered. Beam shape is influenced by a
variety of factors, including the array configuration, pointing direction,
observing frequency and integration time.  A measured shape which is
significantly at variance with expectation is indicative of a failure in
imaging or calibration.

TraP users can predetermine image rejection thresholds for over sampled and
highly elliptical restoring beams as these may be observation or telescope
dependent. All images with restoring beams which are under sampled ($<2$
pixels across the FWHM) are automatically rejected by the TraP.

\subsubsection{Check for proximity to bright radio sources}
\label{sec:components:qc:proximity}

Poor subtraction of bright sources close to the target, during either
`demixing' \citep{vdTol:2007} or subsequent calibration can lead to residual
structures or elevated noise levels in the resultant images. Problems are
typically observed close to the Sun, Jupiter and the so-called ``A-Team'' of
bright extragalactic radio sources \citep{deBruyn:2009}: their extremely high
fluxes may cause issues within target fields up to several tens of degrees
away, depending on the observing configuration. To mitigate these effects, the
TraP rejects images where the angular separation between the target field and
a bright source is less than a user-specified threshold.

\subsection{Source detection and measurement}
\label{sec:components:sf}

The TraP uses a custom-developed source detection and measurement system
(``sourcefinder''). The algorithms implemented are partially based on those
used in SExtractor \citep{Bertin:1996}, but have been extensively re-worked
and extended, most notably to provide least-squares fitting of detected
sources with elliptical Gaussians and a rigorous handling of the correlated
noise properties of radio images. In addition, it provides an implementation
of a false detection rate algorithm \citep{Benjamini:1995, Hopkins:2002}.

In brief, given an input image, the system performs the following procedure:

\begin{enumerate}
  \item{Model and subtract a variable background across the image;}
  \item{Model the RMS noise across the image;}
  \item{Identify islands of contiguous pixels which appear above some multiple
  of the RMS;}
  \item{Decompose multi-component islands into their constituent parts;}
  \item{Perform initial estimation of source parameters;}
  \item{Fit each component with an elliptical Gaussian and return the result.}
\end{enumerate}

This process results in a list of measurements describing the parameters of
each source, including its position, flux density and shape as well as the
significance of the detection: see Table~\ref{tab:sf:results} for details.

More detail on each of the sourcefinder stages is given below. For a thorough
treatment, the reader is referred to \citet{Spreeuw:2010}, while
\citet{Carbone:2015} presents the results of extensive testing of the
sourcefinder system.

\subsubsection{Background and RMS estimation}
\label{sec:components:sf:maps}

The background and noise characteristics of radio astronomy images are
complex. In particular, noise across the image is correlated due to the nature
of the imaging process. Since our source detection procedure relies on
identifying pixels above the RMS noise, careful modelling of the background and
noise maps is essential.

We start by dividing the image into a rectangular grid. The dimensions of the
grid cells are user-specified; they should be chosen such that they are
significantly larger than extended sources visible in the image, but fine
enough to trace background variations.

We estimate the background and RMS in each grid cell through a process of
iterative clipping around the median to reject source pixels. While doing
this, it is important not to bias the result by rejecting source-free pixels,
and to take account of the noise correlation scale.

On each iteration, we label the total number of pixels in a given cell $N$. We
then define the number of independent pixels as $N_\mathrm{indep} = N /
N_\mathrm{dep}$, where $N_\mathrm{dep}$ is the number of pixels per
synthesized beam. We assume that source-free independent pixels follow a
normal distribution, while pixels contaminated by sources do not. We therefore
reject all pixels that fall more than some threshold $n\sigma$ from the
median, where $\sigma$ is the standard deviation of the distribution. The
value of $n$ is chosen such that we will reject, on average, one half of one
normally distributed pixel. That is,
\begin{equation}
N_\mathrm{indep} \times 2\times(1 - F(n\sigma)) = 0.5
\label{eq:cliplimit}
\end{equation}
where $F(n\sigma)$ is the cumulative distribution function for the assumed
normal distribution over the range $[-\infty, n\sigma]$:
\begin{align}
F(n\sigma) &= 0.5 + 0.5 \times \textrm{erf}\left( \frac{n\sigma}{\sigma\sqrt2} \right) \\
           &= 0.5 + 0.5 \times \textrm{erf}\left( \frac{n}{\sqrt2} \right) \\
           &= 0.5 + 0.5 \times \frac{2}{\sqrt\pi}\int_{0}^{n} \mathrm{e}^{-t^2} \mathrm{d}t.
\end{align}
Inverting this, the threshold for clipping is
\begin{equation}
n\sigma = \sigma\sqrt{2}\times\mathrm{erfc}^{-1}\left( \frac{1}{2 N_\mathrm{indep}} \right )
\end{equation}
where $\mathrm{erfc}^{-1}$ is the complementary inverse error function
\citep{Gautschi:1972}.

We estimate the sample variance based on this independent pixel
count\footnote{An alternative approach would be to calculate the pixel
autocorrelation function and the corresponding estimation bias correction
factor \citep{Wolter:1984}; the practical difference is minimal for plausible
values of $N_\mathrm{indep}$.}. That is,
\begin{equation}
\sigma_\mathrm{meas}^2 = \frac{N_\mathrm{indep}}{N_\mathrm{indep} - 1}\left(\overline{x^2} - \overline{x}^2\right)
\end{equation}
where $\sigma_\mathrm{meas}^2$ is the measured sample variance and $x$
represents individual pixel values. However, note that measuring the variance
of a sample which has been clipped at some threshold $T$ causes us to
underestimate the variance as follows:
\begin{equation}
\sigma_\mathrm{meas}^2 = \frac{\frac{1}{\sigma\sqrt{2\pi}} \int_{-T}^{T} x^2 \exp\left(\frac{-x^2}{2\sigma^2}\right) \mathrm{d}x}
                              {\frac{1}{\sigma\sqrt{2\pi}} \int_{-T}^{T} \exp\left(\frac{-x^2}{2\sigma^2}\right) \mathrm{d}x}.
\end{equation}
We invert this to estimate a corrected variance
\begin{equation}
\sigma^2 = \sigma_\mathrm{meas}^2 \frac{ \sqrt{2\pi} \mathrm{erf}\left(T/\sigma\sqrt{2}\right) }
                                       { \sqrt{2\pi} \mathrm{erf}\left(T/\sigma\sqrt{2}\right) - 2T\exp\left(T^2/2\sigma^2\right)/\sigma}
\end{equation}
Note that the ratio, $T/\sigma$, of the clipping limit to the underlying
standard deviation is simply the value of $n$ which was derived following
Eq.~\ref{eq:cliplimit} in the previous iteration.  Finally, following
\citet{Bolch:1968}, we apply a further correction to estimate the standard
deviation, $\sigma$, as
\begin{equation}
\sigma = s \times c_4 = s \times \left(1 - \frac{1}{4N_\mathrm{indep}} - \frac{7}{32N_\mathrm{indep}^2}\right).
\end{equation}
At this point, if any pixel values are more than the calculated $n\sigma$ from
the mean, they are removed from further consideration and a new iteration is
started. Otherwise, the clipping process has completed.

After the pixel values have been clipped, if the field is crowded
($|\textrm{mean}-\textrm{median}|/(\textrm{std.\ dev.}) \ge 0.3$) we take the
background as equal to the median; otherwise, we estimate it as
\begin{equation} \textrm{background} = 2.5 \times \textrm{median} - 1.5 \times
\textrm{mean} \end{equation} following \citeauthor{Bertin:1996}.

Background and RMS maps are then derived by bilinear interpolation of the
background and standard deviation calculated in each grid cell.

\subsubsection{Identification of significant pixel ``islands''}
\label{sec:components:sf:islands}

The background map is subtracted from the input image. The noise map is
multiplied by a user-specified constant (``$n$'') to provide a threshold for
finding sources. Sources are identified where the value of the
background-subtracted input data is larger than the detection threshold map.

Imposing an $n\sigma$ threshold in this way implies that the fraction of
pixels falsely labelled as sources in the image will be smaller than
\begin{equation}
\textrm{erfc}\left(\frac{n}{\sqrt2} \right) = \frac{2}{\sqrt{\pi}} \int_{n/\sqrt2}^\infty \exp\left(-t^2\right) \mathrm{d}t.
\end{equation}

It may be more convenient, however, to quantify the number of false detections
independently of the number of pixels processed.  It is possible to control
the false discovery rate using the algorithm described by
\citet{Benjamini:1995}. This permits the user to specify a maximum number of
``false positives'' (noise peaks erroneously identified as sources), and the
algorithm automatically chooses an appropriate threshold for identifying
sources.

\subsubsection{Deblending}
\label{sec:components:sf:deblend}

After the peaks of pixel islands have been identified, the islands are
expanded to consist of all pixels which are contiguous with the peak pixels
and which are above a further user-specified constant (``$a$'', with $a \le
n$) times the threshold map.

These expanded pixel islands may consist of a single source, but may also
include two or more sources in close proximity. We separate out the components
of these composite sources in a process referred to as ``deblending''
\citep{Bertin:1996,Spreeuw:2010,Hancock:2012}. In the deblending process, we
define a number of sub-thresholds exponentially spaced between the lowest and
highest pixel values in the islands. We iterate through the thresholds from
lowest to highest, at each threshold checking if the island has split into two
(or more) non-contiguous components, each containing a significant fraction of
the integrated flux density, and each with a peak above the detection
threshold $n\sigma$. If these criteria are met, we split the island and regard
it as two separate sources. Both the number of sub-thresholds and the fraction
of the integrated flux density required for a sub-island to be regarded as
significant are user-set parameters.

\subsubsection{Estimating source parameters}

The peak flux density, $P_{\mathrm{max}}$, can be approximated by the value of
the maximum pixel in the island. However, the true source peak will not
coincide with the centre of the pixel. Therefore, we extend the method
described by \citet{Rengelink:1997}, based on the assumption that the true
peak lies at a random position within the pixel. This results in a correction
factor of
\begin{equation}
\iint e^{\ln(2) \left[ \frac{\left( x \cos(\theta) + y \sin(\theta)\right)^2}{m^2} +
                       \frac{\left( y \cos(\theta) - x \sin(\theta)\right)^2}{M^2}
                \right] } \mathrm{d}x \mathrm{d}y
\end{equation}
where $M$, $m$ and $\theta$ are respectively the major and minor axes and the
position angle of the synthesized beam and the integral runs over the pixel.
This correction factor is multiplied by $P_{\mathrm{max}}$ to produce the
output peak flux density.

The total flux density, $F$, is simply the sum of the pixel values
\begin{equation}
F = \sum_{i \in{} S} I_i,
\end{equation}
where $I_i$ is the value of the pixel at position $x_i, y_i$ and $i \in{} S$
indicates all the pixels in a particular island.

The position of the centre of the island is given in pixel coordinates as:
\begin{equation}
\overline{x}, \overline{y} = \frac{\sum_{i \in{} S} I_i x_i}{\sum_{i \in{} S} I_i}, \frac{\sum_{i \in{} S} I_i y_i}{\sum_{i \in{} S} I_i}.
\end{equation}
The position angle of the semi-major axis, measured counter-clockwise from the
$y$-axis, is given by
\begin{equation}
\tan(2\theta{}) = \frac{2 \overline{xy}}{\overline{x^2} - \overline{y^2}}.
\end{equation}
The semi-major ($M$) and semi-minor ($m$) axis lengths are initially estimated
as
\begin{equation}
\begin{Bmatrix}
M^2 \\
m^2
\end{Bmatrix}
= \frac{\overline{x^2} + \overline{y^2}}{2}
\begin{Bmatrix}
+ \\
-
\end{Bmatrix}
\sqrt{\left(\frac{\overline{x^2} - \overline{y^2}}{2}\right)^2 + \overline{xy}^2}.
\end{equation}
These axis lengths are underestimated due to the $a\sigma$ cut at the edge of
the island. They are corrected by multiplying by a factor $\left(1 +
\ln(T/P_{\mathrm{max}})/(P_{\mathrm{max}}/T-1)\right)^{-0.5}$, where $T$ is
the value of the RMS map times the analysis threshold $a$ at the pixel
position of $P_{\mathrm{max}}$.

\subsubsection{Gaussian fitting}
\label{sec:components:sf:gauss}

An elliptical Gaussian is fitted to each island by minimizing the error
function using a modified Levenberg-Marquardt algorithm \citep{More:1977}. By
default, the estimated source parameters calculated above are used as initial
values for fitting, and all parameters are allowed to vary. However, the user
may optionally choose to hold one or more parameters to a fixed,
user-specified value. Typically, this is used to constrain the fitted shape of
point-like sources to that of the restoring beam.

Uncertainties on the parameters $x_0, y_0$ (the fitted pixel position),
$\alpha, \delta$ (the position in celestial coordinates), $\theta_M, \theta_m$
(the lengths of the major and minor fitted axes), $C$ (the fitted peak flux
density), $I$ (the integrated flux density) and $\phi$ (the position angle) are
calculated following \citet{Condon:1997}, \citet{Condon:1998} and
\citet{Hopkins:2003}.  We start by defining a generalized ``signal-to-noise
ratio'' as
\begin{equation}
\rho^2 = \frac{\theta_M \theta_m}{4\theta_B \theta_b}
         \left[1 + \frac{\theta_B^2}{\theta_M^2}\right]^{\alpha_M}
         \left[1 + \frac{\theta_b^2}{\theta_m^2}\right]^{\alpha_m}
         \frac{C^2}{\sigma^2}
\end{equation}
where $\sigma$ is the RMS noise at the location of the source and $\alpha_M,
\alpha_m = (1.5, 1.5)$ for amplitude errors, $(2.5, 0.5)$ for errors on $x$
and $\theta_m$ and $(0.5, 2.5)$ for errors on $y$, $\theta_M$ and $\phi$
\citep{Condon:1997}.

Given the above definitions, we apply the relationships described by
\citet{Condon:1997} and \citet{Hopkins:2003} to obtain
\begin{align}
\frac{2}{\rho^2} &= \frac{\sigma_C^2}{C^2} \\
                 &= 8 \ln2 \frac{\sigma_{y_0}^2}{\theta_M^2} = 8 \ln2 \frac{\sigma_{x_0}^2}{\theta_m^2} \\
                 &= \frac{\sigma_{\theta_M}^2}{\theta_M^2} = \frac{\sigma_{\theta_m}^2}{\theta_m^2} \\
                 &= \frac{\sigma_\phi^2}{2} \frac{(\theta_M^2 - \theta_m^2)^2}{\theta_M^2\theta_m^2} \\
 \sigma_\alpha^2 &= \sigma_{x_0}^2\sin^2\phi + \sigma_{y_0}^2\cos^2\phi \\
 \sigma_\delta^2 &= \sigma_{x_0}^2\cos^2\phi + \sigma_{y_0}^2\sin^2\phi.
\end{align}
Note that, for simplicity, the above assumes that the fitted major axis of the
source aligns with the $y$ axis of the pixel grid. If this is not the case, a
further rotation is required.  Finally, the integrated flux density and its variance
are given by
\begin{align}
                     I &= C \frac{\theta_M\theta_m}{\theta_B\theta_b} \\
\frac{\sigma_I^2}{I^2} &= \frac{\sigma_C^2}{C^2} +
                          \frac{\theta_B\theta_b}{\theta_M\theta_m}\left[
                               \frac{\sigma_{\theta_M}^2}{\theta_M^2} +
                               \frac{\sigma_{\theta_m}^2}{\theta_m^2}
                          \right].
\end{align}

After fitting, the Gaussian restoring beam is deconvolved from the resultant
source parameters using an algorithm derived from that provided by AIPS
\citep{Greisen:2003}, so that the deconvolved shape parameters $\vartheta_M$,
$\vartheta_m$ and $\varphi$ are given by
\begin{align}
\beta^2 &= (\theta_M^2 - \theta_m^2) + (\theta_B^2 - \theta_b^2) \\
        &\phantom{=} -2(\theta_M^2 = \theta_m^2)(\theta_B^2 - \theta_b^2)\cos2(\phi - \Phi) \\
2 \vartheta_M^2 &= (\theta_M^2 + \theta_m^2) - (\theta_B^2 + \theta_b^2) + \beta \\
2 \vartheta_m^2 &= (\theta_M^2 + \theta_m^2) - (\theta_B^2 + \theta_b^2) - \beta \\
\varphi &= \frac{1}{2}\mathrm{atan}\frac{(\theta_M^2-\theta_m^2)\sin2(\phi - \Phi)}
                                        {(\theta_M^2-\theta_m^2)\cos2(\phi - \Phi) - (\theta_B^2 -\theta_b^2)}
           + \Phi.
\end{align}

Following this procedure, the parameters described in
Table~\ref{tab:sf:results} are returned by the sourcefinder and are ready for
insertion into the pipeline database.

\begin{table}
\begin{tabular}{lll}
\toprule
Name                       & Units      & Notes \\
\midrule
Right ascension ($\alpha$) & J2000 deg. & \\
Declination ($\delta$)     & J2000 deg. & \\
Peak flux density          & Jy/beam    & \\
Integrated flux density    & Jy         & \\
Significance               & -          & Peak/RMS \\
Lengths of semi-axes       & arcsec     & \\
Position angle             & deg.       & North through east \\
Error radius               & arcsec     & Absolute error on \\
                           &            & source centroid \\
\bottomrule
\end{tabular}
\caption{Parameters returned by the sourcefinder routines.}
\label{tab:sf:results}
\end{table}

\subsection{Source association}
\label{sec:components:assoc}

Each individual astronomical source detected in a given set of images is
assigned a unique identification in the form of an entry in the ``running
catalogue''.  The running catalogue ties together a series of measurements
made in individual images with aggregated information about the source derived
from those measurements (its position, variability information, etc; see
\S\ref{sec:components:aggregate}). The complete set of measurements associated
with a particular running catalogue entry comprise its lightcurve.

\subsubsection{Association procedure}
\label{sec:components:assoc:proc}

The association procedure adopted is based on \citet{deRuiter:1977},
\citet{Sutherland:1992} and \citet{Rutledge:2000}, and is described in detail
in \citet{Scheers:2011}.  For each measurement, the source association
procedure searches for counterparts in the running catalogue. The algorithm
relies on the de Ruiter radius, the angular distance on the sky between source
$i$ and its potential association counterpart $j$ normalized by the positional
error of both sources. The de Ruiter radius is defined as
\begin{equation}
\label{eq:deruiter}
r_{i,j} = \sqrt{
        \frac{(\alpha_i - \alpha_j)^2 \cos^2 \left((\delta_i + \delta_j) / 2 \right)}{\sigma^2_{\alpha_i} + \sigma^2_{\alpha_j}}
        +
        \frac{(\delta_i -\delta_j)^2}{\sigma^2_{\delta_i} + \sigma^2_{\delta_j}}
        }
\end{equation}
where $\alpha_n$ is the right ascension of source $n$, $\delta_n$ is its
declination, and $\sigma_q$ represents the error on the quantity $q$.

If sources $i$ and $j$ are genuinely associated, their positional differences
will be due to measurement errors, and hence follow a Rayleigh distribution
\citep[e.g.][]{deRuiter:1977}. The probability of source association at $r
\geq \rho$ is then
\begin{equation}
p_r(r \geq \rho) = \int_{r=\rho}^{\infty} r \exp(-r^2/2) \mathrm{d}r = \exp(-\rho^2/2).
\end{equation}

This may be used for determining the search radius, $r_{\mathrm{s}}$, of the
area that will be scanned for possible counterparts: a search radius of
$r_{\mathrm{s}} \leq 3.71$, will miss a factor of $10^{-3}$ of the possible
association counterparts, while $r_{\mathrm{s}} \leq 5.68$ will miss a factor
of $10^{-7}$.

Given the above definition, the source association procedure regards a
particular measurement as being associated with a given running catalogue
source if their positions are no further apart than the semi-major axis of the
restoring beam and the de Ruiter radius is less than a user-specified
threshold. Note that the calculations above only consider repeat measurements
of a single, isolated source. If the TraP is to be used in processing
observations of crowded or extremely transient-rich fields of view, this will
require further consideration of the trade-off in search radius between missed
self-associations, and spurious associations between distinct sources. Making
an optimal choice of search radius for a crowded field will depend on the
precise spatial clustering of sources, an issue which is not investigated
further here.

\subsubsection{Association types}
\label{sec:components:assoc:types}

\begin{figure}
  \begin{subfigure}[]{\columnwidth}
    \centering
    \includegraphics[width=0.9\columnwidth]{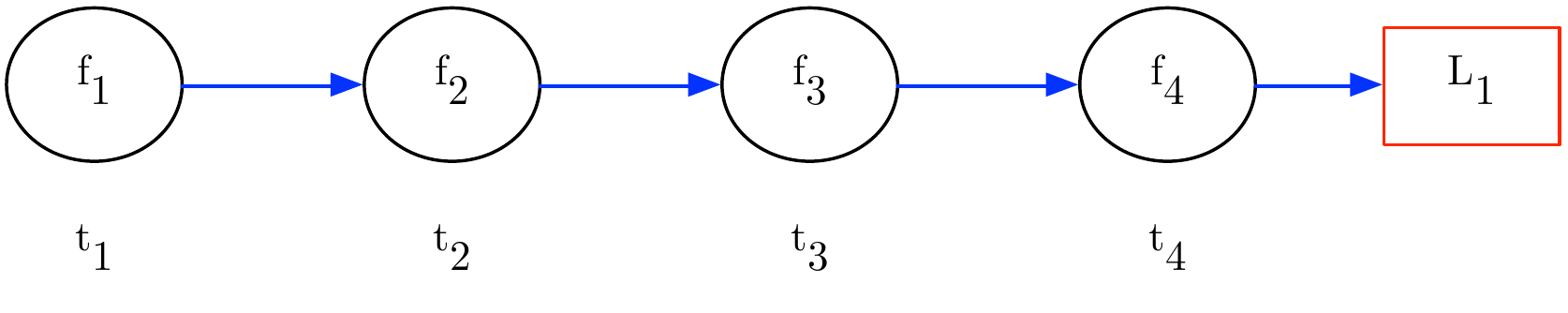}
    \caption{One-to-one.}
    \label{fig:assoc:one2one}
  \end{subfigure}
  \begin{subfigure}[]{\columnwidth}
    \centering
    \includegraphics[width=0.9\columnwidth]{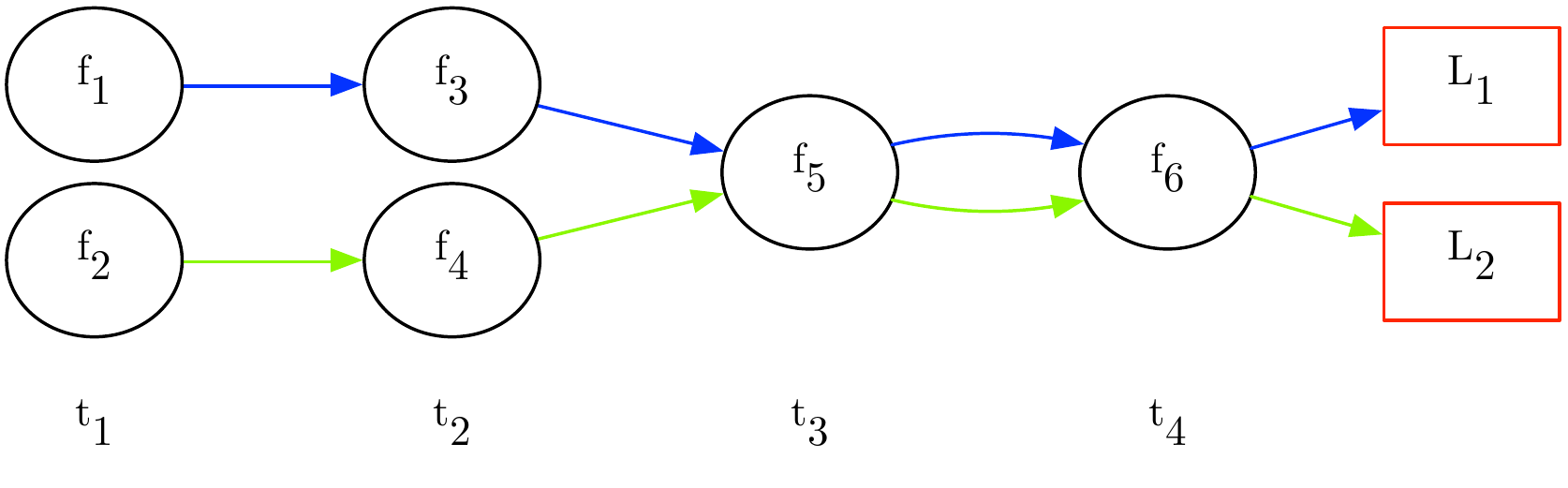}
    \caption{Many-to-one.}
    \label{fig:assoc:many2one}
  \end{subfigure}
  \begin{subfigure}[]{\columnwidth}
    \centering
    \includegraphics[width=0.9\columnwidth]{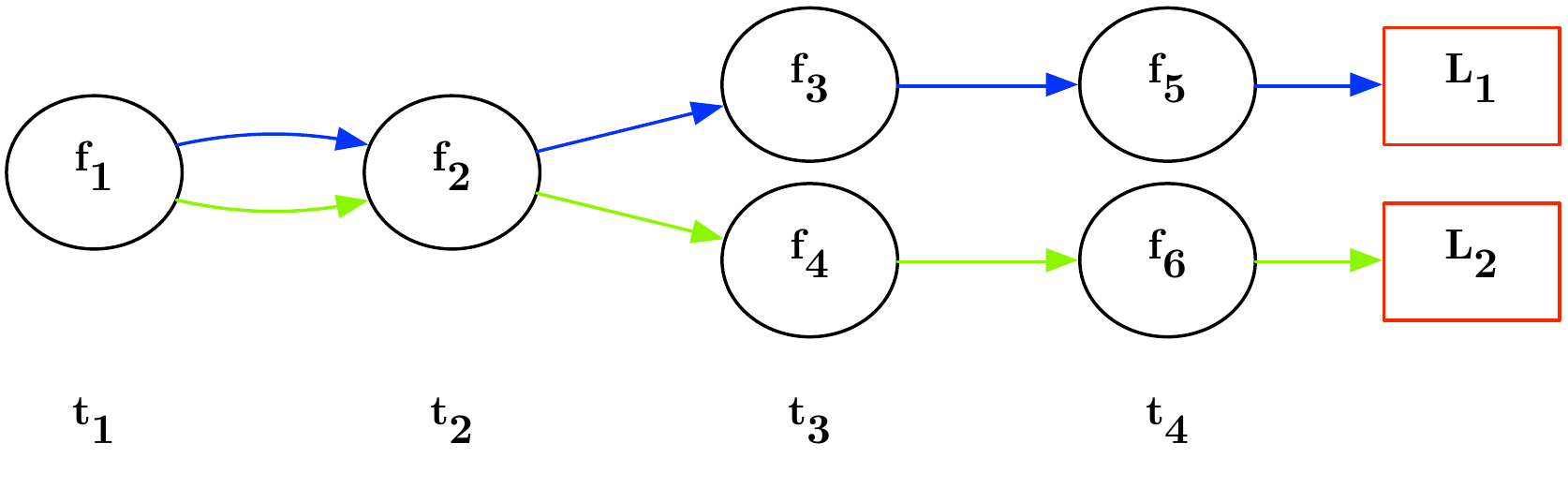}
    \caption{One-to-many.}
    \label{fig:assoc:one2many}
  \end{subfigure}
  \begin{subfigure}[]{\columnwidth}
    \centering
    \includegraphics[width=0.9\columnwidth]{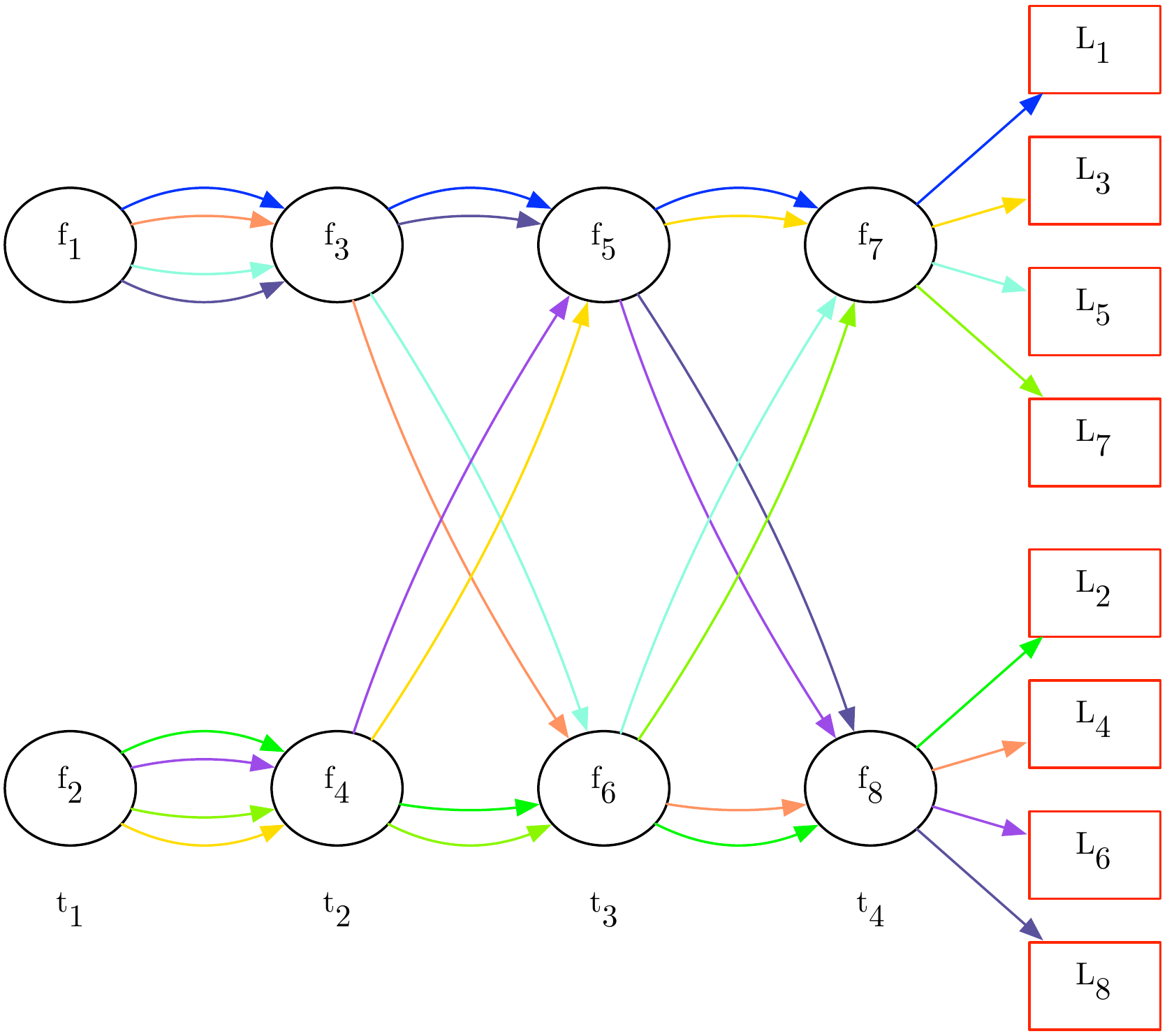}
    \caption{Many-to-many.}
    \label{fig:assoc:many2many}
  \end{subfigure}
  \begin{subfigure}[]{\columnwidth}
    \centering
    \includegraphics[width=0.9\columnwidth]{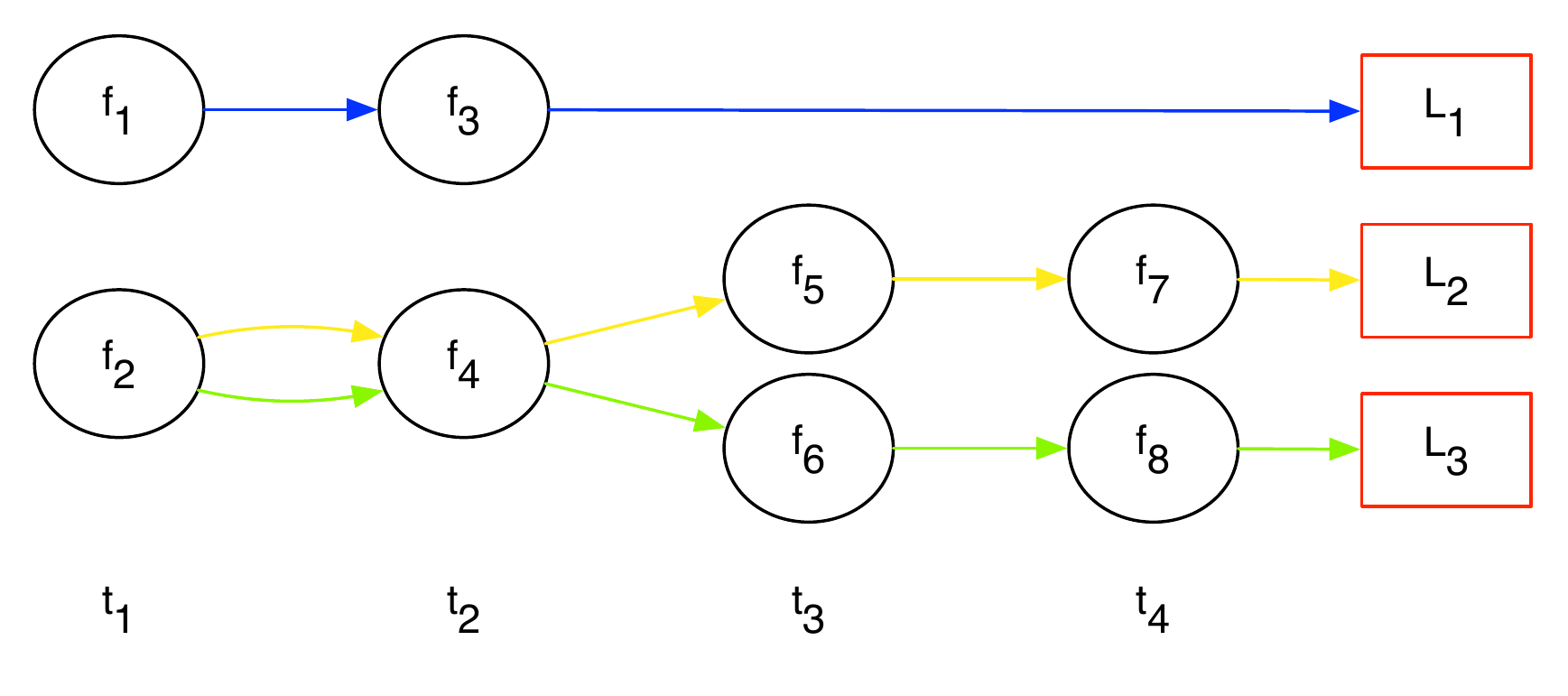}
    \caption{Reduced many-to-many.}
    \label{fig:assoc:many2many-reduced}
  \end{subfigure}
  \caption{Types of source association, corresponding to those described in
  \S\ref{sec:components:assoc:types}. Flux density measurements taken at a
  particular position at time $t_a$ are labelled $f_b$. The association
  procedure knits flux density measurements together between timesteps to form
  lightcurves which are identified with particular running catalogue entries
  identified as $L_c$.}\label{fig:assoc}
\end{figure}

The procedure described above does not guarantee a unique association between
each source measurement and a single running catalogue source. Instead, there
are various possible association topologies, as illustrated in
Fig.~\ref{fig:assoc} and discussed below.

Note that the association is based upon the positions and associated
uncertainties of a particular measurement and the equivalent aggregated
quantities for running catalogue sources; no reference is made to time or
frequency in assessing whether a particular pair is associated. The discussion
below refers to ``time'', but the same considerations apply to association
along a frequency axis. This has the consequence that the order in which data
is associated affects the result, and hence the reproducibility of a
particular analysis. This is discussed in more detail in
\S\ref{sec:components:aggregate:reproduce}.

\paragraph{No association}

The source measurement cannot be associated with any previously catalogued
sources. We regard this as a newly detected source, and create a catalogue
entry for it.

\paragraph{One-to-one}

The flux density measurement is unambiguously associated with a single running
catalogue entry, as shown in Fig.~\ref{fig:assoc:one2one}. The average flux
density of the source $L_1$ is $\overline{f_{1\cdots{}4}}$.

\paragraph{Many-to-one}

Many-to-one associations exist when two or more running catalogue sources
satisfy the association criteria for a given measurement. We record this in
the database as multiple separate one-to-one associations. This is illustrated
in Fig.~\ref{fig:assoc:many2one}: at times $t_1$ and $t_2$ distinct
lightcurves, $L_1$ and $L_2$ are being tracked. However, at $t_3$ a single
source is detected which can be associated with both of these lightcurves.
This could happen, for example, if the third observation was made at a lower
angular resolution.

Note that the single flux density measurements $f_5$ and $f_6$ are now
independently included in two separate sources: $L_1$ has average flux density
$\overline{f_{1,3,5,6}}$ and $L_2$ has average flux density
$\overline{f_{2,4,5,6}}$.  The total brightness reached by summing all
catalogued source flux densities has been artificially increased. Since all
individual measurements are stored, it is possible for users to correct for
this in their analysis according to their particular requirements. Future
versions of the TraP may support the division of flux density from a single
measurement among multiple sources (\S\ref{sec:future}).

\paragraph{One-to-many}

In the reverse of the previous case, a single running catalogue source can be
associated with more than one source measurement from a given image. This is
handled by splitting the catalogue source into two independent sources, $L_1$
with average flux density $\overline{f_{1,2,3,5}}$ and $L_2$ with average flux
density $\overline{f_{1,2,4,6}}$ as shown in Fig.~\ref{fig:assoc:one2many}. As
in the many-to-one case, some source measurements are included in multiple
lightcurves, artificially increasing the total brightness of the catalogue.

\paragraph{Many-to-many}

A many-to-many association occurs when more than one running catalogue source
can be associated with more than one extracted source. If the procedures
described above were applied, every possible combination of catalogue sources
and new measurements would result in a new lightcurve: the database complexity
would increase quadratically, and the situation rapidly becomes untenable.
This is illustrated in Fig.~\ref{fig:assoc:many2many}. To prevent this,
many-to-many associations are reduced to one-to-one or one-to-many
associations by choosing only the source pairs with the smallest de Ruiter
radii. By applying this procedure, the association topology is reduced to a
simpler case such as that shown in Fig.~\ref{fig:assoc:many2many-reduced}:
rather than the eight separate lightcurves produced by ``pure'' many-to-many
association, we are left with just three.

\subsection{Null-detection handling}
\label{sec:components:nulldet}

We use the term ``null-detection'' to describe a source which was expected to
be observed in some particular image---it had been observed in previous images
of the same field with a flux density above our detection threshold for this
image---but which was not detected by the source finding procedure
(\S\ref{sec:components:sf}).

After the association step (\S\ref{sec:components:assoc}), we build a list of
all sources which:

\begin{itemize}

  \item{Exist in the list running catalogue of known sources, having been
  observed (in any frequency band) at an earlier timestep;}

  \item{Were not associated with any source detected in the current image.}

\end{itemize}

For each of these null detections, we use a modification of the sourcefinding
procedure: the same techniques are employed for background and noise
estimation and source measurement as previously described, but, rather than
being based on islands of pixels above some threshold, the peak of measurement
is constrained to be fixed to the catalogued position of the null detection.
No other additional constraints are placed on the measurement.

After the measurement has been made, it is stored as a one-to-one
association with the appropriate running catalogue entry; the measurement is
marked as being due to the null detection procedure.

In the current release of the TraP, once a running catalogue entry has been
created, forced fitting at that location will always be attempted in the
absence of a matched blind-detection. Ultimately, an accumulation of marginal
single-epoch false detections due to noise will cause a very large number of
irrelevant forced fits to be performed and stored. This may be mitigated by
expiring these marginal detections from the list of positions to be measured
if they are not re-observed after a certain period of time or number of
revisits. Automatic expiry according to user-defined criteria will be included
in a later release of TraP.

\subsection{Monitoring list sources}
\label{sec:components:monitoring}

We anticipate that there may be some sources which are important to particular
science cases which may not always be detected by the default sourcefinding
procedures employed by the TraP. It is therefore possible for the end-user to
supply a ``monitoring list'' of positions at which a measurement will always
be made, regardless of the results of the sourcefinding step. The current
version of the TraP assumes that input images have been correctly registered
when making these measurements: it makes no attempt to correct for astrometric
shift.

For each image which covers the location of a position on the monitoring list,
a measurement is taken at its location. The same procedure is used as for null
detections (\S\ref{sec:components:nulldet}): a modified version of the
algorithms described in \S\ref{sec:components:sf} which hold the position of
the measurement constant.

For each monitored position, a running catalogue source is generated which
contains only a chain of one-to-one associations of measurements at the
user-specified position. Sources monitored in this way are not included in the
general association procedure described in \S\ref{sec:components:assoc:proc}.

\subsection{Aggregate source properties}
\label{sec:components:aggregate}

For each entry in the running catalogue, we now have at least one individual
measurement (corresponding to the current image) and potentially others
corresponding to other images (presumably representing observations at
different times or frequencies). We combine these measurements to estimate the
true source properties as follows.

\subsubsection{Mean values}
\label{sec:components:aggregate:mean}

For each property of a given source we store both the arithmetic mean and a
weighted mean.  For a series of  measurements of property $x$, we denote the
arithmetic mean of $x$ as $\overline{x}$.  We define the weight of a
measurement of property $x$ as
\begin{equation}
\label{eq:weight}
w_x = 1 / \sigma_x^2
\end{equation}
where $\sigma_{x}$ is the uncertainty on that measurement. The weighted mean
of $N$ such measurements is then
\begin{equation}
\xi_{x} = \frac{\sum_{i=1}^{N} w_{x_i} x_i}{\sum_{i=1}^{N} w_{x_i}}.
\label{eq:weightedmean}
\end{equation}

Using these definitions, for each source we calculate both the arithmetic and
weighted means of:

\begin{itemize}
  \item{The source right ascension (J2000 degrees);}
  \item{The source declination (J2000 degrees).}
\end{itemize}

For each frequency band for each source, we calculate both the arithmetic and
weighted means of:

\begin{itemize}
  \item{The peak flux density of the source in this band (Jy/beam);}
  \item{The integrated flux density of the source in this band (Jy).}
\end{itemize}

\subsubsection{Variability metrics}
\label{sec:components:aggregate:variability}

After source association, each running catalogue entry corresponds to a
multi-frequency lightcurve of a particular source. We search for transient and
variable sources by examining the properties of these lightcurves.

For each source, we label a measurement of its flux density at a particular
frequency, $\nu$, as $I_{\nu}$ with uncertainty $\sigma_\nu$. Based on these
quantities, and using the same notation as above, we define the flux density
coefficient of variation\footnote{This quantity is occasionally referred to as
the `modulation index' in astronomical literature
\citep[e.g.][]{Narayan:1992,Gaensler:2000,Jenet:2003,Bell:2014}. The present
authors prefer `coefficient of variation' due to its more widespread use
\citep{McKay:1932,Hendricks:1936,Lande:1977,Freund:2010} and because it avoids
any possible confusion with other fields \citep[e.g.][]{Whitaker:1996}.} over
$N$ measurements as the ratio of the mean flux density to the sample standard
deviation $s$, thus:
\begin{equation}
V_\nu = \frac{s}{\overline{I_\nu}}
      = \frac{1}{\overline{I_\nu}} \sqrt{\frac{N}{N-1}\left(\overline{{I_\nu}^2}-\overline{I_\nu}^2\right)}.
\label{eq:V_nu}
\end{equation}

Using the same definition of weight $w_\nu = 1/\sigma_\nu^2$ as above, we can
also express the significance of the flux density variability. We make use of
the familiar reduced-$\chi^2$ expression in conjunction with the weighted mean
flux density, $\xi_{I_\nu}$:
\begin{equation}
\label{eq:eta_nu}
\eta_\nu = \frac{1}{N-1} \sum_{i=1}^N \frac{\left(I_{\nu, i} - \xi_{I_\nu} \right)^2}{\sigma_{\nu, i}^2}.
\end{equation}

For a given $\eta_\nu$, the probability that we can reject the null
hypothesis---that the source under consideration is not variable---is given by
\begin{equation}
P_{\mathrm{variable}} = 1 - \int_{{\eta_\nu}'=\eta_\nu}^{\infty}p\left({\eta_\nu}',N-1\right)\mathrm{d}{\eta_\nu}'.
\end{equation}
where $p(x, n)$ is the $\chi^2$ probability density function for $x$ over $n$
degrees of freedom \citep[see, for example,][]{Kesteven:1977}.

$V_\nu$ and $\eta_\nu$ are calculated and stored for every lightcurve whenever
a new source measurement is added. Since variability metrics are stored per
association, we can track how the variability parameters of a source have
changed with time. This is particularly useful in the case of those sources
which have shown evolution in their behaviour over time.

It is worth noting that it would be relatively straightforward to extend the
TraP to support the calculation and storage of other variability metrics
beyond the two described above. It is expected that extended testing and
experience in processing data from various sources will guide future
development in this area.

\subsubsection{Reproducibility of results}
\label{sec:components:aggregate:reproduce}

Reproducibility of pipeline results is of paramount importance: the end user
should be confident that repeatedly running the TraP on a given dataset with
the same configuration should always produce the same results. This has
important consequences for the association and aggregation procedures. For
example, consider a particular running catalogue source $R$ and two source
measurements, $M_1$ and $M_2$, taken from different images. If $M_1$ is
inserted first, it is associated with $R$. On association, a new aggregate
position for $R$ is calculated (which may or may not be consistent with
association with $M_2$). On the other hand, if $M_2$ is inserted first, the
resulting aggregate position for $R$ is \textit{not} consistent with
association with $M_1$. In short, the order in which the images are processed
influences the association procedure, and hence changes the outputs.

In order to mitigate this effect, the TraP only guarantees reproducibility of
output if the input is in monotonically increasing order of time. If two or
more images with different frequency or Stokes parameters but the same
timestamp are processed, the TraP will automatically sort them along these
axes before processing. This is not, in general, possible along the time axis,
which is potentially unbounded.

\section{Data products}
\label{sec:products}

After all the stages described in \S\ref{sec:components} have been completed
for a given image cube, the core pipeline loop is complete. The complete
running catalogue, together with ancillary data describing the images which
have been processed, pipeline configuration and other metadata, is now stored
in a database, the structure of which is described in detail in
\S\ref{sec:implement:db}. At this point, the pipeline may start processing the
next image cube. Simultaneously, the contents of that database may be used to
support scientific analysis and alert generation.

\subsection{Identifying transient events}
\label{sec:products:alerts}

As described in \S\ref{sec:overview}, we distinguish between newly detected
and variable sources. Both are scientifically significant and may form the
basis for a ``new transient'' alert. Depending on context and science goal,
these alerts may simply result in a particular source being brought to the
attention of the pipeline end user, or they may be distributed more widely to
the community. The technical mechanism used for alert dissemination is
described in \S\ref{sec:implement:voevent}.

\subsubsection{New detections}
\label{sec:products:alerts:new}

As described in \S\ref{sec:components:assoc:types}, a measurement which cannot
be associated with a previously catalogued source is regarded as a new
detection. Such a new detection might correspond to a new transient source
appearing, but it could also simply mean that this area of sky had not
previously been surveyed to a depth adequate to detect this source.

In order to distinguish between these possibilities, the TraP keeps track of
the fields-of-view and sensitivities of all images it processes. When a new
source is detected, its flux density is compared against the recorded
sensitivities of images covering the same area to see if it could have been
detected previously. If so, it may be regarded as a new transient.

In practice, and as described in \S\ref{sec:components:sf:maps}, noise (and
hence sensitivity) is not constant across any given image. It is possible that
a particular source could have been detected if it fell in a low-noise area of
a previous image, but not if it fell in a high-noise area. We therefore record
both the highest and lowest RMS values recorded in each previous image
($\sigma_{\mathrm{max},i,\nu}$ and $\sigma_{\mathrm{min},i,\nu}$ in image $i$
at frequency $\nu$) as well as the detection threshold ($n_{i,\nu}$) used when
processing that image. For a flux density measurement $I_{i,\nu}$,
we regard it as a marginal transient candidate if
\begin{equation}
\sigma_{\mathrm{max},j,\nu} \times (n_{j,\nu} + M) > I_{i,\nu} \ge \sigma_{\mathrm{min},j,\nu} \times (n_{j,\nu} + M).
\end{equation}
where $M$ is some user specified margin, applied to prevent stable sources
with flux densities approximately equal to the detection threshold from being
misidentified as transients. This marginal category would include both genuine but
faint transients, and steady state sources which change in significance as
noise varies between images. In the case that
\begin{equation}
I_{i,\nu} \ge \sigma_{\mathrm{max},j,\nu} \times (n_{j,\nu} + M) \quad \forall j < i
\end{equation}
we regard the source as likely to be a new transient.

\subsubsection{Variable sources}

As per \S\ref{sec:components:aggregate:variability}, three separate
variability metrics---the flux density coefficient of variation $V_\nu$, the
significance of the variability $\eta_\nu$, and the probability of variability
$P_{\mathrm{variable}}$---are stored whenever a new source association is
made. We can therefore search the catalogue for sources which meet the
following criteria:

\begin{itemize}
  \item{$V_\nu$ is above a user-specified threshold;}
  \item{$\eta_\nu$ is above a user-specified threshold;}
  \item{$P_{\mathrm{variable}}$ is above a user-specified threshold;}
  \item{The number of source measurements in the lightcurve is above a
  user-specified threshold.}
\end{itemize}

Choosing appropriate values for these thresholds is a matter of user
configuration and will depend on the details of the science case as well as
the particular dataset under investigation. Section~\ref{sec:test} gives an
overview of possible considerations, while \citet{Rowlinson:2015} presents a
detailed consideration of how optimal parameters might be chosen.

\subsection{Database interface}
\label{sec:products:db}

While automatic routines may be used to scan the database for transients as it
is constructed using the methods described in \S\ref{sec:products:alerts}, it
is likely that many scientific discoveries will be based on expert analysis of
the database contents. Here, we describe the systems by which this is made
available to end users.

\subsubsection{Direct database queries}
\label{sec:products:db:sql}

The core database engine is one of various off-the-shelf database management
systems, as discussed in detail in \S\ref{sec:implement:db}. Given appropriate
permissions\footnote{Database permissions are controlled by the administrators
of a particular TraP installation; it is possible for them to grant
permissions to both query and modify the data to arbitrary users if
required.}, the database may be queried directly using SQL\footnote{Structured
Query Language, the de-facto standard for interacting with relational database
systems.}. Expert users may write SQL scripts or use the command line
interface to the database system to manually construct complex queries
selecting exactly the information they require.

\subsubsection{Web interface}
\label{sec:products:db:banana}

%
%

While the ability to query the database for arbitrary information which
answers a specific science question is undeniably powerful, it requires a
significant investment of time on the part of the end user to become fluent in
the relevant technologies. Further, it is often convenient to have an
at-a-glance summary of the contents of the database and the important results.
For this reason, the TKSP has developed \textit{Banana}, a web-based interface
to the database. Banana enables the user to conveniently examine the contents
of the database, viewing details (including cut-out images) of all source
measurements, plotting lightcurves of all sources in the running catalogue,
selecting potential transients based on their variability metrics, and so on.

Banana is open-source (it is released under a Modified BSD
license\footnote{\url{http://opensource.org/licenses/BSD-3-Clause}}) and is
developed and released by the TKSP independently of the TraP. It is freely
available for download\footnote{\url{https://github.com/transientskp/banana}}.

\subsubsection{High volume archival data-mining}
\label{sec:products:db:archive}

For modest input dataset sizes (thousands of images, tens or hundreds of
sources per image), the total volume of data stored is modest: on the order
of, perhaps, gigabytes. However, as per \S\ref{sec:lofar:rsm}, the TraP
ultimately aims to support long term operation of the LOFAR RSM, which will be
capable of producing thousands of images per hour. It is also a stated aim of
the project to make the lightcurve archive available to the community as a
legacy resource. Efforts are currently underway to both develop database
systems capable of handling this volume of data \citep{Scheers:2011,
Scheers:2014}, and a batch query system akin, for example, to CasJobs
\citep{OMullane:2005} is under consideration. Ultimately, we also hope to make
data available using a Virtual Observatory-compliant interface. However, the
currently-available IVOA model for time series data \citep{Graham:2014} is an
intentionally minimal interim solution; we prefer to wait for more mature
standards to become available \citep[e.g.][]{McDowell:2015} before proceeding.

\section{Implementation}
\label{sec:implement}

In this section, we turn our attention to the underlying technology which
implements the workflow described in sections
\ref{sec:overview}~and~\ref{sec:components}.

\subsection{Architecture}
\label{sec:implement:architecture}

The TraP is structured as a series of pipeline ``steps'', each of which
performs a logically discrete operation in turn. These steps, and the
relationship between them, are shown graphically in Fig.~\ref{fig:trap}.

The operations carried out by the TraP naturally split into those which
involve directly interacting with image data (such as loading data from disk
or finding and measuring sources within it) and those which operate on
measurements derived from the data (source association or aggregate
calculation).

Interacting with images and performing measurements upon them is most
effectively accomplished by bespoke software packages developed by members of
the TKSP which directly encode the required algorithms. We have developed a
series of such packages using the Python programming
language\footnote{\url{http://www.python.org/}}. The choice of Python for this
task, together with a discussion of the approach taken, is motivated in
\S\ref{sec:implement:python}.

The derived data which results from source measurements is highly structured.
It can be efficiently stored using a relational database system
\citep{Codd:1970}. The contribution of the database goes beyond mere storage,
however: by performing calculations within the database itself it possible to
operate on the entire corpus of data efficiently and with minimal overhead due
to data transport. In this way, the database becomes the core computational
engine of the TraP. The design and structure of the database is described in
\S\ref{sec:implement:db}.

A control layer sits above the core scientific logic as defined in Python code
and the database. This control layer defines the structure of the
pipeline---effectively connecting the components together in the correct
order---as well as providing utility services such as parallelization and task
distribution, which we describe in \S\ref{sec:implement:parallel}.

Finally, and in addition to the pipeline routines described, the TraP offers
the option to save a copy of all the pixel data processed to a separate
document-oriented database for later use by the Banana web interface
(\S\ref{sec:products:db:banana}). This is described in
\S\ref{sec:implement:mongodb}.

\subsection{Python}
\label{sec:implement:python}

Python is the primary programming language used in the TraP. We consider
Python to be the default choice for astronomical software development where
performance is not the critical consideration in the near to intermediate
future. It provides a flexible and expressive language together with a wide
ecosystem of scientific and other libraries, and it is easily extensible using
code written in lower-level languages where maximum performance is
required\footnote{We have experimented with writing portions of the TraP in
Cython (\url{http://www.cython.org/}) for performance reasons with some
success, but no Cython code is shipped with the current release.}.
Furthermore, thanks to projects like IPython \citep{Perez:2007} and AstroPy
\citep{Astropy:2013}, Python is also increasingly finding a role in the daily
workflow of many astronomers as an interactive data analysis toolbox. Although
we do not directly use these tools in the TraP, this familiarity then lowers
the barrier to entry on larger projects as the novice coder becomes more
proficient, potentially widening the pool of future maintainers and
contributors to open-sourced scientific codes.

Although we have had great success using Python, a significant downside is
that, as a dynamically typed language, there is a risk of run-time type
errors. We have countered this by adopting a strongly test-focused development
style and building an extensive suite of TraP unit tests. This is a topic we
return to in \S\ref{sec:dev:unittest}.

\subsection{Parallelization and distribution}
\label{sec:implement:parallel}

Some operations which are carried out by the TraP can be performed
concurrently on multiple datasets. For example, the initial source finding and
measurement step (\S\ref{sec:components:sf}) can be performed on many
independent images simultaneously without changing its results. We can exploit
this intrinsic parallelism to obtain the best possible run-time performance by
distributing processing across multiple CPU cores (or even distinct machines)
and scheduling as many operations to run concurrently as is possible.

Of course, the performance improvement which may be achieved in this way is
limited: some operations (such as source association,
\S\ref{sec:components:assoc}) cannot run concurrently (we rely on data being
processed in a particular order to ensure reproducibility;
\S\ref{sec:components:aggregate:reproduce}); there is an intrinsic ordering of
pipeline steps (it is impossible to perform source association before source
measurement is complete); and, for reasons of reproducibility, we mandate that
all steps relating to images corresponding to a particular observation time
are complete before a subsequent observation time can start processing
(\S\ref{sec:components:aggregate:reproduce}). These intrinsically sequential
parts of the processing limit its overall performance \citep{Amdahl:1967}.

We have implemented the TraP in such a way that the definition of the
underlying algorithms is independent of the method used to start tasks and
collect results. In this way, it is possible to insert different task
scheduling back-ends which support different parallelization and distribution
techniques. Three are currently supported by the TraP:

\begin{itemize}

  \item{The \textit{serial} back-end runs tasks sequentially using a single
  Python interpreter. Using the standard Python interpreter\footnote{The
  reference ``CPython'' implementation.} means that all (non-database)
  processing takes place in series on a single CPU core.}

  \item{The \textit{multiproc} back-end uses the \texttt{multiprocessing}
  package\footnote{\texttt{multiprocessing} is part of the Python standard
  library.} to schedule jobs on multiple CPUs within a single machine
  concurrently.}

  \item{The \textit{celery} back end uses
  Celery\footnote{\url{http://www.celeryproject.org/}}, an asynchronous task
  distribution system, to marshal the distribution of concurrent TraP tasks
  across a cluster of multiple machines.}

\end{itemize}

The end user may select which back-end to use when invoking the TraP from the
command line.

Note that the \texttt{celery} system does not arrange for data to be
transmitted across the cluster. If, for example, it is used to distribute a
source finding step across multiple images, it is required that each machine
have access to the particular images which it is to process (perhaps on its
local disk or on shared storage). This is a convenient match to the imaging
process on the LOFAR cluster, which deposits image data on the cluster node
which was responsible for creating it. We have, nevertheless, prototyped an
image transmission system which is better integrated with the TraP, but it is
not included in the current release.

Many-core exploitation works well for the `embarrassingly parallel' problem of
sourcefinding across many different images (corresponding to different
frequencies and pointings). However, the bottleneck for processing a timestep
then becomes the database operations, placing stringent performance
requirements on the combination of query complexity and database back-end used,
as covered in Section~\ref{sec:implement:db}.

\subsection{Database}
\label{sec:implement:db}

All source measurements, together with metadata describing the images from
which they were taken, are stored and processed in a relational database. The
contents of the database is itself one of the core products of the TraP
(\S\ref{sec:overview:products}), while queries run over the database are an
intrinsic part of regular pipeline processing. Only a single instance of the
TraP may write to a given database during processing (but the database may be
accessed by arbitrary read-only queries).  However, both of the supported
database management systems, as described in \S\ref{sec:implement:db:systems},
provide for multiple isolated databases being hosted within a single system,
so supporting multiple pipeline instances is straightforward. This could be
used, for example, to support multiple independent monitoring campaigns.

\subsubsection{Database management systems}
\label{sec:implement:db:systems}

The TraP is developed and tested using two relational database management
systems (RDBMS): MonetDB\footnote{\url{http://www.monetdb.org/}} and
PostgreSQL\footnote{\url{http://www.postgresql.org/}}.

MonetDB is an ongoing project to build a database with exceptional performance
and scalability based on research into data organization and query
optimization \citep{Idreos:2012}. The potential performance benefits of
MonetDB are impressive, particularly when considering the ultimate data
volumes expected from the RSM (\S\ref{sec:lofar:rsm}) and, later, from the
SKA\@. However, its development is driven by fundamental database research and
scientific user groups, and it may occasionally perform in unexpected or
undesirable ways. We therefore also verify the correct operation of the TraP
and, where necessary, provide ``production grade'' deployments using
PostgreSQL, which has a long pedigree as an industry-standard database
management system.

The TraP pipeline code and the Banana web interface send queries to the
database using SQL\@. Although SQL is a standardized language\footnote{ISO/IEC
9075-1:2011}, there is significant variation in its implementation between
different database vendors: code that is written and tested against one
database may unexpectedly fail when run on another system. Therefore, while
the code in the TraP is designed to be standards compliant and database vendor
agnostic, it is occasionally necessary to add special cases to work around
different SQL dialects. To accommodate this the TraP provides a simple
templating system for SQL queries. For example, we can accommodate both the
PostgreSQL and MonetDB syntaxes for defining a function within the database as
follows:
\begin{verbatim}
{% ifdb monetdb %}
CREATE PROCEDURE BuildFrequencyBands()
{% endifdb %}

{% ifdb postgresql %}
CREATE OR REPLACE FUNCTION BuildFrequencyBands()
RETURNS void
AS $$
{% endifdb %}

BEGIN
  -- Definition elided
END;
\end{verbatim}
In this way it is easy to extend the TraP's database support to encompass
other systems if required.

\subsubsection{Database structure}

\begin{figure*}
  \begin{center}
  \includegraphics[width=289pt]{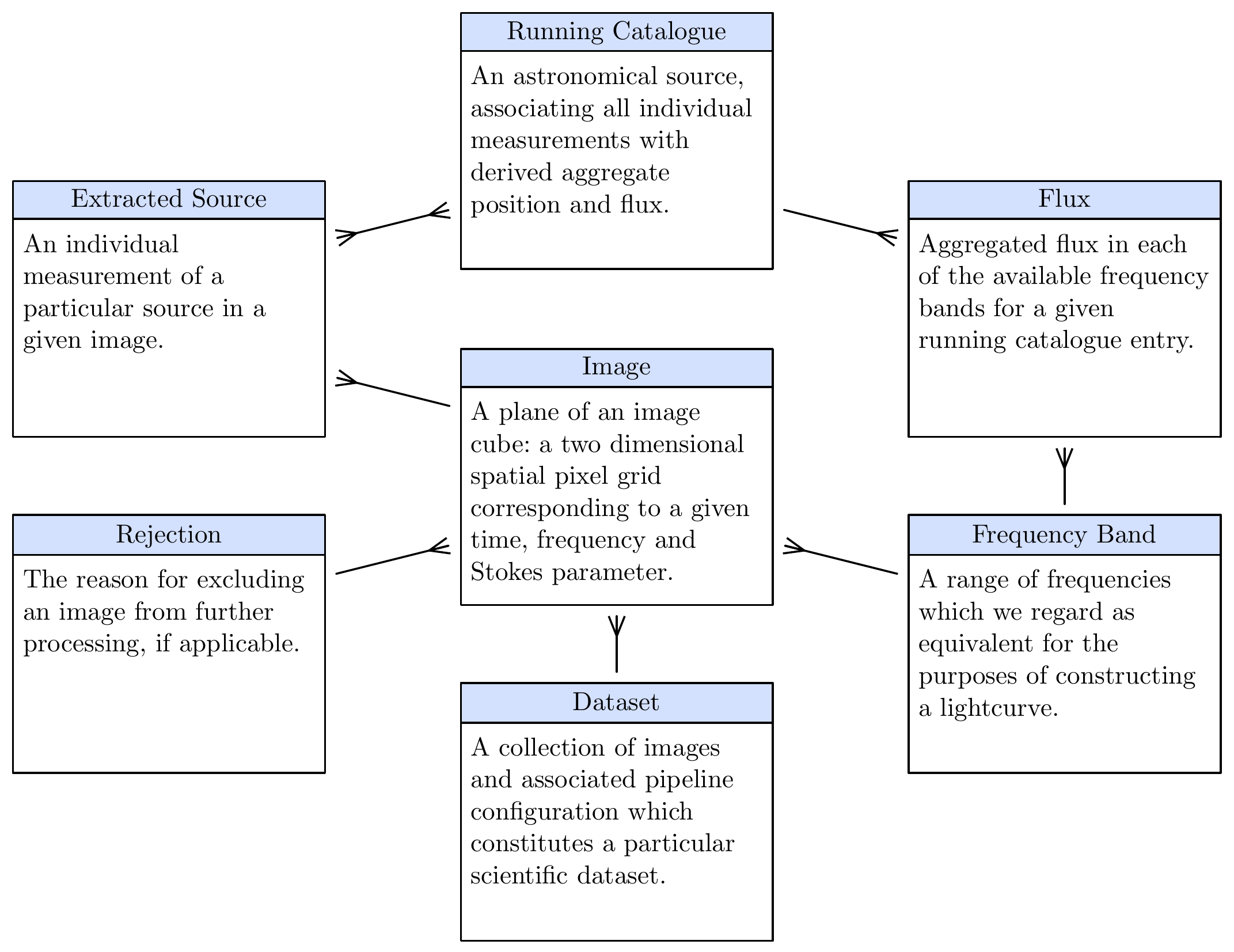}
  \end{center}

  \caption{Simplified version of the TraP database structure. Each box
  represents a database table, while lines represent key-based
  cross-references. A single line refers to a unique key, while a split line
  indicates that a many-to-one relationship is possible. For example, an
  \code{Image} is a member of a single \code{Dataset}, but a \code{Dataset}
  may contain many \code{Image}s.}

  \label{fig:simpleschema}
\end{figure*}

The structure of the current version of the TraP database is elaborate,
consisting of many separate tables, some with tens of individual columns,
complex relationships between them, and a variety of stored procedures. It is
described in detail in the TraP
Handbook\footnote{\url{http://docs.transientskp.org}}. A simplified version is
shown in Fig.~\ref{fig:simpleschema}.

It is expected that, as TraP development progresses, the database will evolve
to meet new requirements. For this reason, the database is versioned, and the
TraP does not support mixing results across versions. The current codebase
does not offer explicit support for ``schema migrations'' (converting a
database from one version to another without losing data), although versions
with this functionality have been prototyped, and off-the-shelf schema
migration tools\footnote{For example, Alembic
\url{http://alembic.readthedocs.org/}.} may be applicable.

It is worth noting, for performance reasons, the database is not fully
normalized \citep{Kent:1983}.

\subsubsection{Positional database queries}
\label{sec:implement:db:positional}

Perhaps the most common operation required of the database, both during
pipeline processing (e.g. \S\ref{sec:components:assoc}) and during later
analysis is the ``cone search'': finding all objects of a particular type
(such as source measurements, running catalogue entries) within a given search
radius of a particular position. Since this operation is so common, we give it
special consideration.

Given a particular target $(\alpha_\tau, \delta_\tau)$, search radius $r$ and list
of positions $(\alpha_1, \delta_1), (\alpha_2, \delta_2), ... (\alpha_n,
\delta_n)$, the simplest approach to finding all the positions which fall
within the radius of the target is to iterate down the list calculating for
each one the great-circle distance
\begin{equation}
r_{\tau, n} = \arccos(\sin\delta_\tau\sin\delta_n + \cos\delta_\tau\cos\delta_n\cos|\alpha_\tau-\alpha_n|)
\label{eq:greatcircle}
\end{equation}
and selecting only those sources for which $r_{\tau, n} \leq r$. While
conceptually simple, this involves multiple calculations for every source to
be checked: it is prohibitively computationally expensive given a large source
list.

Since this simple approach is impractical, we adopt an approach based on that
described by \citet{Gray:2006}\footnote{Note that a variety of alternative
approaches were considered, such as HTM \citep{Szalay:2005} and Q3C
\citep{Koposov:2006}; practical experience and compatibility with the database
management systems informed the approach taken.}. This involves a hierarchical
approach, filtering the list of candidate sources first on declination then on
right ascension before selecting candidates based on a Cartesian dot product.

First, when any new source measurement is inserted into the database, or when
the weighted mean position (\S\ref{sec:components:aggregate:mean}) of a
catalogue source is updated, we calculate a corresponding position on the unit
sphere:

\begin{align}
x &= \cos\delta \cos\alpha \label{eq:cart:x} \\
y &= \cos\delta \sin\alpha \label{eq:cart:y} \\
z &= \sin\delta \label{eq:cart:z}
\end{align}

At the same time, we define a function $\mathtt{zone}(\delta)$ which maps a
declination to a particular ``zone'', corresponding to a strip of the sky.
Zones must increase monotonically with declination: $\mathtt{zone}(\delta_m)$
includes all declinations falling between $\mathtt{zone}(\delta_{m-1})$ and
$\mathtt{zone}(\delta_{m+1})$. In the current version of the TraP we use the
largest integer smaller than $\delta$ as the zone, thus:
\begin{equation}
\mathtt{zone}(\delta) = \lfloor \delta \rfloor.
\end{equation}

However, this definition is flexible: given the constraint above, future
versions could adopt a zone definition with a more fine-grained resolution or
variable zone heights (e.g.\ chosen to provide zones of uniform area).

The Cartesian coordinates and the zone are stored in the database and are
henceforth available for each running catalogue source and source measurement
with no further run-time calculation.

It is next necessary to describe the `inflation' of angular distances in right
ascension with declination. For example, at a declination of \degrees{0}, a
circular region with radius $\theta$ centred on right ascension $\alpha$
includes RAs in the range $[\alpha-\theta, \alpha+\theta]$, whereas at a
declination of \degrees{90} it covers the complete circle. Following
\citeauthor{Gray:2006}, we define the function $\mathtt{alpha}(\theta,
\delta)$ as
\begin{equation}
\mathtt{alpha}(\theta, \delta) = \left\{ \begin{array}{ll}
    \arctan \frac{\sin \theta}{\sqrt{\cos(\delta-\theta) \cos(\delta+\theta)}} & \mbox{if $|\delta|+\theta < 90$}; \\
    180 & \mbox{otherwise.} \end{array} \right.
\end{equation}
In general, such a circle at arbitrary $\alpha, \delta$ can be said to cover
the range $[\alpha-\mathtt{alpha}(\theta, \delta),
\alpha+\mathtt{alpha}(\theta, \delta)]$. $\mathtt{alpha}(\theta, \delta)$ is
implemented as a stored procedure directly in the database so that it can be
calculated for arbitrary $\theta$ and $\delta$ with minimal overhead.

These definitions made, we are now able to quickly filter the list positions
to be searched. First, we calculate the maximum and minimum zones in which
targets may be found, rejecting all those targets for which
$\mathtt{zone}(\delta_n)$ lies outside the range
$[\mathtt{zone}(\delta_{\tau-r}), \mathtt{zone}(\delta_{\tau+r})]$. Then we
reject all targets for which $\alpha_n$ lies outside the range
$[\alpha_\tau-\mathtt{alpha}(r, \delta), \alpha_\tau+\mathtt{alpha}(r,
\delta)]$. By ensuring that the database is appropriately indexed on
$\mathtt{zone}(\delta_n)$ and $\alpha$ this filtering can be done extremely
fast.

The above filtering reduces the potentially large list of positions to match
to a much more manageable size. For each position, we now check whether it
lies within the required angular distance of the target. Rather than
calculating the great circle distance (Eq.~\ref{eq:greatcircle}) it is more
efficient to use a scalar product based on the Cartesian positions calculated
according to Eqs.~\ref{eq:cart:x}--\ref{eq:cart:z}. Thus we check for:
\begin{equation}
x_\tau \cdot x_n + y_\tau \cdot y_n + z_\tau \cdot z_n \leq \cos{r}
\end{equation}

Since $(x_n, y_n, z_n)$ for each candidate is already stored in the database,
the total amount of computation (and hence running time) is kept to a minimum.

The above procedure fails in the case of a discontinuity in RA: at the
meridian, we jump from \degrees{359} to \degrees{0}, breaking the check for
$\alpha_n$ lying within a given range. To work around this, if a source
association query crosses the meridian, we rotate the RAs of the relevant
sources by \degrees{180} to avoid the discontinuity, perform the association
as normal, and then rotate the results back to the original orientation.

\subsubsection{Iteratively updating aggregate quantities}
\label{sec:implement:db:update}

As per \S\ref{sec:components:aggregate}, we store weighted mean positions,
flux densities and variability indices for all sources in the database. When a
new measurement is appended to a particular running catalogue entry, it would
be possible to re-calculate these quantities from scratch by averaging over
all the existing information about that source in the database. This would
clearly be inefficient, though. Instead, we update these quantities
iteratively.

The arithmetic mean of some property $x$ over $N$ measurements is
\begin{equation}
\overline{x}_N = \frac{1}{N} \sum_{i=1}^N x_i
\end{equation}
where $x_i$ is the $i^\mathrm{th}$ measurement of $x$. When a further
measurement of $x$ is taken, updating the mean iteratively is straightforward:
\begin{equation}
\overline{x}_{N+1} = \frac{N\overline{x}_{N} + x_{N+1}}{N+1}.
\end{equation}

As per \S\ref{sec:components:aggregate:mean}, we also calculate the weighted
mean defined as in Eq.~\ref{eq:weightedmean}.  The weight of measurement $x_i$
is $w_{x_i}$; the sum of all weight over $N$ measurements is
\begin{equation}
W_{x_N} = \sum_{i=1}^N w_{x_i}.
\end{equation}

Given $\xi_{x_N}$ and $W_{x_N}$ it is then possible to express the weighted
mean after $N+1$ measurements as

\begin{align}
\xi_{x_{N+1}} =& \frac{W_{x_N}\xi_{x_N} + w_{x_{N+1}}x_{N+1}}{W_{x_N} + w_{x_{N+1}}} \\
              =& \frac{N\overline{w}_{x_N}\xi_{x_N} + w_{x_N+1}x_{N+1}}{N\overline{w}_{x_N} + w_{x_{N+1}}}.
\end{align}

Therefore, by storing the number of measurements ($N$) and the average weight
($\overline{w}_{x_n}$) in addition to the weighted mean, we can iteratively
update the mean and weighted mean source properties as new measurements are
added to the database without revisiting all previous measurements.

As measurements are associated with catalogue sources, we keep track not only
of the mean parameters but also of the evolving variability parameters, $V_\nu$
and $\eta_\nu$, as described in \S\ref{sec:components:aggregate:variability}.

Updating $V_\nu$ is straightforward. We store both the mean and the mean
square flux density per band, $\overline{I_\nu}$ and $\overline{I_\nu^2}$ and
update them using the procedure described above. They can then be directly
used to calculate $V_\nu$ based on Eq.~\ref{eq:V_nu}.

To handle $\eta_\nu$, we substitute Eq.~\ref{eq:weight} into
Eq.~\ref{eq:eta_nu} to get
\begin{align}
\eta_\nu =& \frac{N}{N-1} \left(\frac{1}{N} \sum_{i=1}^{N} w_{\nu, i} (I_{\nu, i} - \xi_{I_\nu})^2\right) \\
         =& \frac{N}{N-1} \left(\frac{1}{N} \sum_{i=1}^{N} w_{\nu, i} (I_{\nu, i}^2 - 2I_{\nu, i}\xi_{I_\nu} + \xi_{I_\nu}^2) \right) \\
         =& \frac{N}{N-1} \left(\overline{w_{\nu} I_{\nu}^2} - 2\xi_{I_\nu} \overline{w_\nu I_\nu} +\xi_{I_\nu}^2 \overline{w_\nu}\right) \\
         =& \frac{N}{N-1} \left(\overline{w_\nu\,{I_\nu}^2}-\frac{\overline{w_\nu\,I_\nu}^2}{\overline{w_\nu}}\right)
\end{align}
where we use the definition of $\xi_{I_\nu}$ from Eq.~\ref{eq:weightedmean}. This
quantity can be calculated directly from the aggregates stored in the database.

Every time a new association is stored in the database, the values of the
variability parameters calculated at the time of association are stored together
with it. In this way, it is possible to query the database for the variability
parameters corresponding to any point in the history of a particular running
catalogue entry.

\subsection{Pixel store}
\label{sec:implement:mongodb}

Images are not created during the operation of the TraP; in general, therefore,
we regard their storage as outwith the scope of TraP operations. However, it is
often convenient to maintain easy access to image data which has been processed.
This enables end users who are analysing the TraP results to quickly cross-check
them with a visual inspection of the image data. Indeed, tools such as Banana
(\S\ref{sec:products:db:banana}) can over-plot details of sources identified by
the TraP on the image data.

In normal operation, the TraP reads images from the filesystem attached to
whichever machine (or machines) upon which it is executing. Often, that
filesystem is not intended as long-term image storage, but is rather a temporary
resting place on whatever compute system is being used for analysis. Further, it
may not always be desirable (for security or management reasons) for the
ultimate scientific user of the TraP to have access to the systems upon which
the pipeline runs. Finally, it is simply more convenient to aggregate images for
display in one location, rather than have Banana or other tools search for them
on diverse filesystems.

For these reasons, the TraP can optionally insert a copy of all pixel data it
processes to a centralized store. The term ``pixel data'' is used deliberately:
rather than storing complete image cubes, with full metadata, images are
reduced to a lowest common denominator form consisting of just a pixel grid and
coordinate system stored in FITS format. This enables a convenient and uniform
interface by which data may be accessed for display, but does not amount to a
comprehensive archive of the images.

The pixel storage used by the TraP is implemented as a
MongoDB\footnote{\url{https://www.mongodb.org/}} database. MongoDB is a
``document-oriented'' database, which makes it easy to simply store and retrieve
large ``blobs'' of binary data (such as our pixels) using a simple key-value
look-up scheme.

Pixel data may be saved to the MongoDB database by the data accessor
(\S\ref{sec:components:accessors}) when it is first loaded from disk. A URL
identifying the location of the corresponding pixels is then stored in the
\texttt{Image} table of the main TraP database (Fig.~\ref{fig:simpleschema}).

\subsection{Dissemination of transient notifications}
\label{sec:implement:voevent}

After an event has been selected as scientifically noteworthy, information
about it must be rapidly distributed. In general, notifications will be sent
to the community at large, although it is possible that certain events may
only be shared with selected partners.

Currently, the rate of transients being announced by LOFAR is low, but we
anticipate it increasing in the future \citep{Fender:2014, Stewart:2014}.
Looking further ahead to facilities like SKA \citep{Dewdney:2010} and LSST
\citep{Ivezic:2014}, it is reasonable to expect that millions of transients
may be announced every day.  Furthermore, rapid turn-around time for follow-up
observations is often necessary.  Therefore, we regard it as imperative that,
as far as is possible, transient alerts can be generated, transmitted,
received and acted upon without human intervention. This makes possible the
development of the automatic systems that will be required to handle the
upcoming transient deluge \citep[see, for example,][]{Staley:2013}.

With the above considerations in mind, we have standardized upon the VOEvent
\citep{Seaman:2011} format developed by the International Virtual Observatory
Alliance (IVOA) for describing transients detected by LOFAR\@. VOEvent
provides a standardized, machine-readable way of describing a celestial event
with the implication that timely follow-up is of interest. VOEvent provides
mechanisms for describing:

\begin{itemize}

  \item{The facility and/or observer responsible for publishing the
  notification packet;}

  \item{A description of the event observed;}

  \item{Where and when the observations where made;}

  \item{Instrument specific information describing how the data was
  collected; and}

  \item{A scientific assessment of the event, which may be used to motivate
  the request for follow-up.}

\end{itemize}

All of this information is presented in an XML document which can be
conveniently manipulated by computer, but it may also be accompanied by
plain text descriptions for human consumption.

The flexibility of this format is such that early LOFAR transient
notifications can be simple (a position, a timestamp, a frequency and a flux
density measurement, for example), and, as our understanding of both the
instrumentation and the low frequency radio sky improves, the event
descriptions can become increasingly elaborate and include detailed
classification information and scientific assessment.

The VOEvent standard does not specify a means by which VOEvents should be
transmitted from originator to recipient. However, ongoing work in the IVOA
and the transient astronomy community has developed a transportation protocol
\citep{Allan:2009} and an early version of a worldwide distribution network
\citep{Williams:2012}. The TKSP team has developed
Comet\footnote{\url{http://comet.transientskp.org/}} \citep{Swinbank:2014}, an
open-source implementation of this transportation system, and will use that to
publish VOEvents to the distribution network. A prototype of a similar system
is currently being used to notify the robotic pt5m
telescope\footnote{\url{https://sites.google.com/site/point5metre/}} of
observations by AMI-LA.

\section{Development methods}
\label{sec:dev}

The TraP is a large and complex project: it consists of some tens of thousands
of lines of code, written in Python and SQL, which are very different
languages; it ingests image data from a variety of different sources; it
interacts with two different types of database; and it is developed, tested
and supported by a heterogeneous team of software developers and academic
astronomers spread across multiple different institutions. Ensuring the
delivery of reliable software which produces scientifically valid results
under these circumstances requires a rigorous development methodology.

\subsection{Planning and issue tracking}

TraP releases are made at the cadence of a few per year. This provides a
compromise between deploying new and upgraded features to end users as rapidly
as possible, and providing a stable base which users can trust to provide
consistent results from day to day while they work on a particular science
project.

Releases alternate between ``science'' and ``technical'' focuses. The
science-focused releases aim to deploy new and upgraded scientific analysis
capabilities. Technically focused releases concentrate on consolidation of the
codebase and introducing new technology, without changing the capabilities
available to end users.

Goals for a release are defined through a series of ``issues'' targeted to a
particular milestone in an issue tracker\footnote{For most of the lifetime of
the TraP to date, this was Redmine, \url{http:///www.redmine.org/}. We have
recently switched to GitHub Issues, \url{http://www.github.com/}, for better
integration with our version control system.}. During the development cycle a
daily test build is made available for commissioners.  In light of development
experience and results from the test build, the issues targeted for the
milestone may be revised, and new issues may be added. When all the issues
targeted for that milestone have been addressed, a release occurs and the
cycle repeats.

\subsection{Code repository and version control}

The most fundamental tool in developing and maintaining a large codebase is a
version control system. This is essential to maintain a list of changes to the
code, including information about who changed what, when, and what the
rationale was. We use the version control system to develop and test multiple
variations of the TraP in parallel; to isolate and revert errors introduced to
the code; and to enable the painless integration of code developed by
different and geographically separate developers.

The TraP makes use of Git\footnote{\url{http://www.git-scm.com/}}, with a
central repository currently hosted on
GitHub\footnote{\url{http://www.github.com/}}. Our experience has been that
software developers are quick to adapt to working with Git, but that its
complexity can be off-putting to those coming from a more purely scientific
background. We have organized training sessions and workshops in order to
mitigate this.

\subsection{Code review}

In order to ensure that all code entering the codebase is of high quality,
and to ensure that there is no single part of the codebase which
is understood by only one developer, we require that all code contributions are
reviewed by a team member other than their original author before they are added
to the TraP. This process is managed using GitHub's ``pull request'' interface.

The overhead introduced by this review step is not negligible: the reviewer must
often invest considerable time to become familiar with the code being reviewed,
and sometimes a lengthy discussion between the original author and the reviewer
can result. Furthermore, it can occasionally be frustrating for the author to
wait for a reviewer to become available during busy times.

Despite these downsides, though, the review process has been successful: since
it was instituted, the quality and reliability of the TraP codebase has
increased markedly, and the entire development team has better insight into
all parts of the pipeline rather than just their own particular
specialization.

\subsection{Testing and continuous integration}
\label{sec:dev:unittest}

Testing is fundamental to the development of any software system.  Mistakes
are inevitable, and, in a large and complex codebase, predicting all possible
effects of even simple changes becomes challenging.  This is particularly the
case when development takes place using a dynamic language such as Python:
with no compile-time checking for type or even syntax, it is easy for errors
to slip by without being noticed.

The TraP codebase is rigorously and automatically tested. At time of writing,
the test suite consists of some 347 individual test cases, with three times
that number of individual assertions contained within them. Test cases cover
everything from `unit' testing (checking that individual functions and
procedures perform as expected when provided with both normal and
extraordinary input conditions) to large scale `integration' tests which
validate the scientific results produced by large sections of the TraP on
given input data. All new code must pass all of these tests (or,
alternatively, explain why the test suite should be changed) before it is
accepted by the code review process. Furthermore, all submissions are expected
to come with their own set of tests which demonstrate their correctness.

Our testing infrastructure is based upon the \texttt{unittest} module provided
as part of Python's standard library and the
Jenkins\footnote{\url{http://jenkins-ci.org/}} continuous integration system.

\subsection{Documentation}

Documentation is provided both for the end-user astronomer who needs to
understand how to process their data and interpret the results, and for the
expert user or developer who is extending the TraP to address their particular
use case.

The TraP is documented using the Sphinx\footnote{\url{http://sphinx-doc.org/}}
documentation system.  This can both automatically generate interface
documentation from the TraP's Python code while also incorporating
hand-written material giving a more complete description of the code along
with tutorial-style documentation. As part of the code review process,
reviewers are expected to check that code not only functions properly and is
well tested, but also that, if appropriate, it is accompanied by appropriate
additions or alterations to the documentation.

The documentation for all released versions of the TraP as well as the latest
developmental version is available from the project
website\footnote{\url{http://docs.transientskp.org/}}.

\section{Integration testing}
\label{sec:test}

As described in \S\ref{sec:dev:unittest}, the TraP codebase is well covered by
an extensive test suite which tests individual components and their
interactions when provided with a variety of different inputs, based on both
synthetic and archival data. Further, individual subsystems and the algorithms
they implement have undergone extensive testing both in the published
literature and in regular use. For example, \citet[chapter 4]{Scheers:2011}
describes how the source association routines were applied to cross-match the
VLSS \citep[VLA Low Frequency Sky Survey;][]{Cohen:2007}, WENSS
\citep[Westerbork Northern Sky Survey;][]{Rengelink:1997} and NVSS \citep[NRAO
VLA Sky Survey;][]{Condon:1998} catalogues. This functionality forms the basis
of the LOFAR Global Sky Model database \citep{vanHaarlem:2013}. Similarly,
\citet[chapter 3]{Spreeuw:2010} describes an elaborate series of statistical
tests on the sourcefinder, which are expanded upon by \citet{Carbone:2015}.
Results from the sourcefinder were also submitted to the ASKAP/EMU\footnote{EMU
is ASKAP's Evolutionary Map of the Universe Survey Science Project} Source
Finding Data Challenge \citep{Hopkins:2015}; the final results of this
exercise have not yet been published, but preliminary indications are that the
TraP code has performed to a high standard.

Although the individual components of the TraP are well tested, it is useful
to consider an integration test, which demonstrates the operation of the TraP
as a coherent whole and provides an indication as to how the results may be
interpreted. It is stressed that this section serves primarily as an
illustration of a pipeline run under strictly controlled circumstances: we do
not attempt to account for complex or unexpected behaviour of astronomical
sources, as this can best be considered by comparing the source behaviour to
the documented sourcefinder capabilities, database sources association
behaviour, variability metrics, etc. It is worth noting that a companion
paper, \citet{Rowlinson:2015}, expands upon the techniques presented here to
establish strategies to determine optimal TraP configuration for a given
dataset given expected source and image characteristics, while early science
results derived from pre-release versions of the TraP are now becoming
available \citep{Carbone:2015a}.

\subsection{Simulation procedure}
\label{sec:test:sim}

Simulated monochromatic lightcurves representing single-epoch transients
observed at a frequency $\nu$ were generated.  Each lightcurve consisted of 20
flux density measurements, with the flux density recorded for measurement $i$,
$I_{\nu, i}$, given by:
\begin{equation}
I_{\nu, i} = \left\{ \begin{array}{ll}
      I_\mathrm{transient} & \mbox{if $i = 8$};\\
      I_\mathrm{quiescent} & \mbox{otherwise}.\end{array} \right.
\end{equation}

The transient flux density, $I_\mathrm{transient}$ was varied over the range [5,
95]\,Jy in steps of 5\,Jy. The quiescent flux density, $I_\mathrm{quiescent}$, was
varied over the range [0\,Jy, $I_\mathrm{transient}$) using the same step
size. In this way, a total of 190 lightcurves were generated.

For each lightcurve, a set of 20 images representing LOFAR observations of the
transient was simulated. In order to closely mimic genuine LOFAR observations,
the simulation developed was based closely on the structure of existing LOFAR
visibility data. We started with visibility data obtained as part of LOFAR's
Multifrequency Snapshot Sky Survey \citep[MSSS;][]{Heald:2014}. The data
consisted of 20 observations of the field of 3C\,295 ($14^h11^m20.6^s$,
$+52^\circ12'21''$) made between 03:00Z and 08:00Z on December 24, 2011. Each
observation had an integration time of 11 minutes, and was followed by a 4
minute re-pointing time. Observation configurations were identical, covering
2\,MHz of bandwidth, divided among 10 subbands and centred on 54\,MHz.

We generated a model sky for the area being observed by selecting all sources
from the VLSS catalogue \citep{Cohen:2007} which fall within 8$^\circ$ of the
pointing centre and which are above a limiting flux density of 1\,Jy. Spectral
indices for these sources were generated by comparing the VLSS flux densities
with those reported in WENSS \citep{Rengelink:1997} and NVSS
\citep{Condon:1998}, then used to extrapolate the source flux density to
54\,MHz. Where a VLSS source had no counterpart in the other catalogues, a
spectral index of -0.7 was assumed.

The simulation procedure for each image was:

\begin{enumerate}

  \item{All subbands were averaged to produce a single channel with a width of
  1\,MHz;}

  \item{The stored visibilities were replaced with randomly generated
  Gaussian noise at a level chosen to match the System Equivalent Flux Density
  of the instrument \citep{vanHaarlem:2013};}

  \item{The appropriate transient flux density, based on the lightcurve being
  processed and the image number, was appended to the model sky at position
  $14^h20^m00.0^s$, $+52^\circ00'00.0''$ (\degrees{1.34} from 3C\,295);}

  \item{BBS \citep{Loose:2008}, the standard tool use for calibrating LOFAR
  data, was used to simulate model visibilities and add them to the data based
  on a user-supplied model sky;}

  \item{The data was calibrated and imaged as usual. The transient source was
  included in the model sky used for calibration. Images were generated
  with a radius of 6$^\circ$.}

\end{enumerate}

The resulting images had an RMS noise level around 0.5\,Jy/beam. Simulated
sources with $I_\mathrm{quiescent} \geq 5$ are therefore detected at around
$10\sigma$ or higher in their quiescent states; sources with
$I_\mathrm{quiescent} = 0$ are detected when the transient turns on in image
8.

\subsection{Predicted results}

\begin{figure*}

  \begin{subfigure}[b]{0.48275862\textwidth}
    \includegraphics[width=\columnwidth]{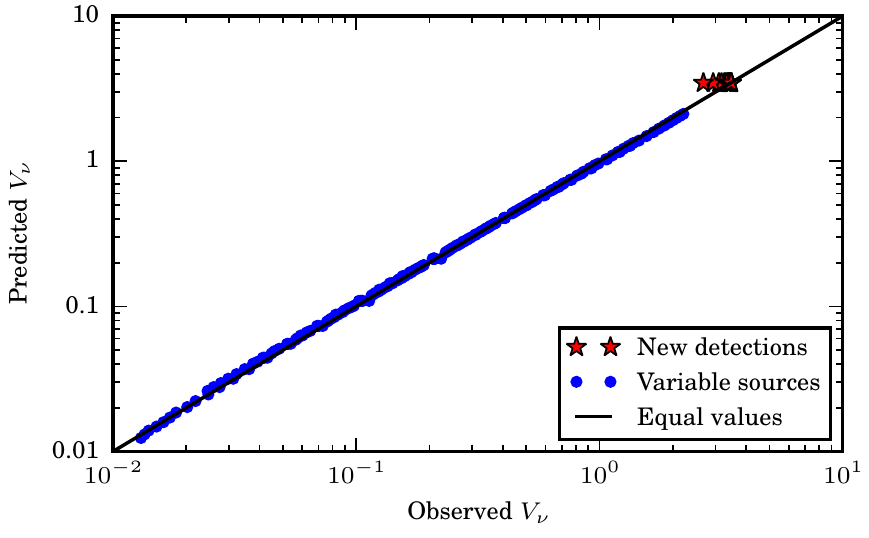}
    \caption{Flux density coefficient of variation ($V_\nu$).}
    \label{fig:predVobs:v}
  \end{subfigure}
  \hfill
  \begin{subfigure}[b]{0.48275862\textwidth}
    \includegraphics[width=\columnwidth]{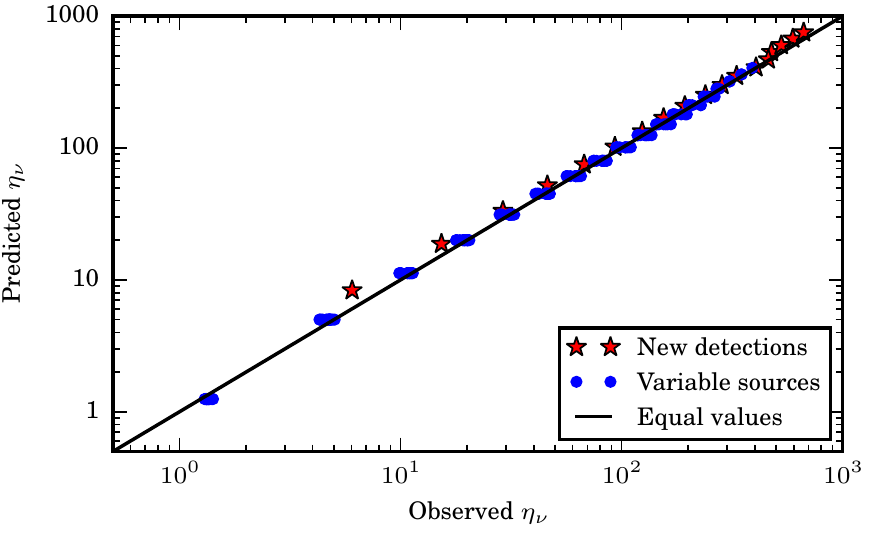}
    \caption{Significance of the flux density variability ($\eta_\nu$).}
    \label{fig:predVobs:eta}
  \end{subfigure}

  \caption{Comparison of predicted and measured variability parameters for
  simulated transients.}
  \label{fig:predVobs}

\end{figure*}

The flux density coefficient of variation, $V_\nu$, and the significance of
the variability, $\eta_\nu$, as described in
\S\ref{sec:components:aggregate:variability} were calculated independently of
the pipeline machinery for each of the 190 lightcurves described in the
previous section. Note that the calculation of these metrics depends not only
on the raw simulated values, as described in the previous section, but also on
the configuration of the pipeline run. In particular:

\begin{itemize}

  \item{When calculating $\eta_\nu$ (Eq.~\ref{eq:eta_nu}), we assigned an
  equal weight (equal to the reciprocal of the average error across all flux
  density measurements) to each data point.}

  \item{Only those flux density measurements recorded in or after the image of
  first detection are included in the variability metric calculation; as per
  \S\ref{sec:overview}, measurements from earlier images are not available
  during pipeline processing.}

\end{itemize}

In this way, we were able to predict the variability metrics which the TraP
should calculate for each source, and determine in advance which ones ought to
be identified as transients for a given TraP configuration.

\subsection{Transients pipeline results}

Each dataset of snapshot images (190 datasets, 1 for each transient) was run
through TraP. A near-default pipeline configuration was used: the quality
control system set to not reject any images and the shape of all point sources
was constrained to be equal to that of the clean beam. The variability metrics
corresponding to the final snapshot for each source were extracted from the
database at the end of each pipeline run. Fig.~\ref{fig:predVobs} shows the
predicted values for each of the transient sources in comparison to the
measured value by the transient pipeline.  The scatter in this figure is due
to three factors:

\begin{itemize}

  \item{The simulation process generates sources on a noisy background, and this
  noise impacts on the results produced by the sourcefinder;}

  \item{When a source is not detected by the initial sourcefinding step, the
  TraP's null-detection procedure (\S\ref{sec:components:nulldet}) will force
  a constrained fit to its position and record a (likely non-zero) flux
  density. The prediction procedure, by contrast, assumes a flux density of
  0\,Jy;}

  \item{The predicted values were calculated assuming an equal weighting for
  all flux density points, whereas the TraP assigns each point an independent
  weight.}

\end{itemize}

Taking these expected deviations into account, the predicted and measured
values are consistent with each other. In particular, we tend to predict
higher variability indices than are measured for faint sources, as our
prediction procedure takes no account of image noise, assigns an equal weight
to all measurements, and assigns a flux density of 0\,Jy to non-detections.

As per \S\ref{sec:products:alerts}, we distinguish between two classes of
transients: \textit{new detections} and \textit{variable sources}.  In this
section we describe targeted tests which confirm that both mechanisms are
performing as expected.

Throughout this section, we refer to ``true positive'' (TP) detections when a
transient source is correctly identified as such; ``false positive'' (FP)
detections when a non-transient source is incorrectly labelled as transient, and
``false negative'' (FN) detections when a transient source is incorrectly
labelled as non-transient. We then define

\begin{align}
{\rm Precision} =& \frac{{\rm TP}}{{\rm TP}+{\rm FP}} \quad \mathrm{and} \label{eq:precision} \\
{\rm Recall} =& \frac{{\rm TP}}{{\rm TP}+{\rm FN}}. \label{eq:recall}
\end{align}

Following these definitions, the \textit{precision} of the result is the
fraction of the total number of detections which are correct and the
\textit{recall} is the fraction of the total number of transients which were
correctly identified. The best possible TraP performance is obtained by
maximizing both the precision and the recall.

\subsubsection{New detections}

\begin{figure*}
  \begin{subfigure}[b]{0.48275862\textwidth}
    \includegraphics[width=\textwidth]{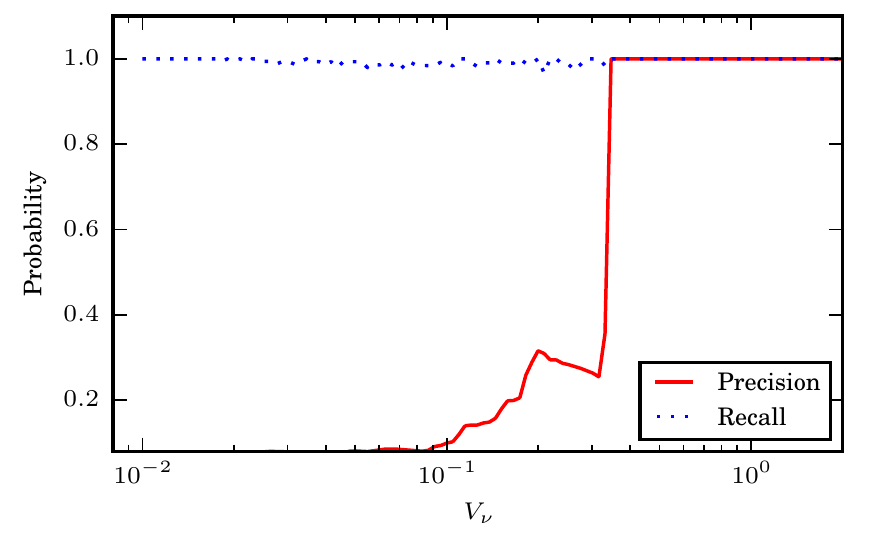}
    \caption{Probabilities as a function of the threshold applied to the flux
    density coefficient of variation ($V_\nu$).}
    \label{fig:FP_properties:v}
  \end{subfigure}
  \hfill
  \begin{subfigure}[b]{0.48275862\textwidth}
    \includegraphics[width=\textwidth]{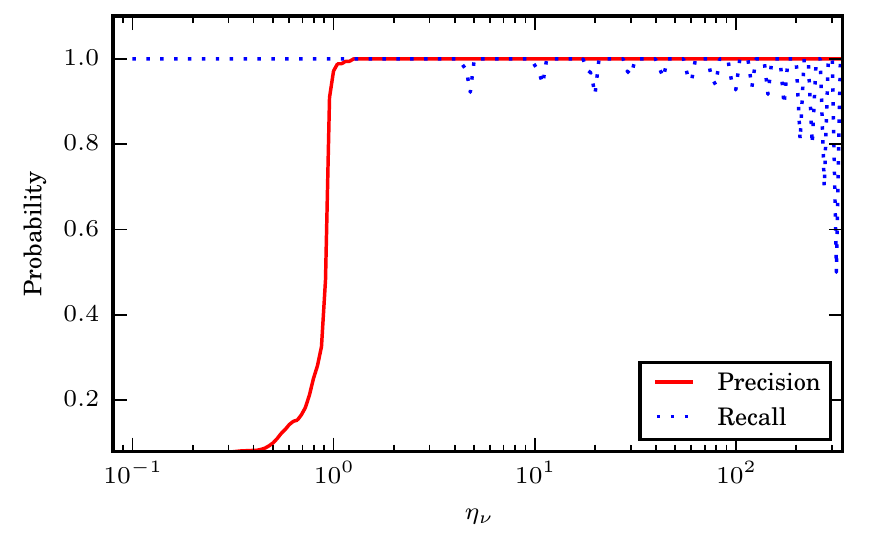}
    \caption{Probabilities as a function of the threshold applied to the
    significance of the flux density variability ($\eta_\nu$).}
    \label{fig:FP_properties:eta}
  \end{subfigure}

  \caption{The precision and recall probabilities as functions of the
  triggering thresholds for the variability metrics $V_\nu$ and $\eta_\nu$.}
  \label{fig:FP_properties}

\end{figure*}

We selected all sources from the database that were initially detected in any
image except the first. Following the procedure described in
\S\ref{sec:products:alerts:new}, and using a margin of $3\sigma$, we
classified them as either not transient (i.e.\ below previous detection
threshold), marginally transient (above previous detection threshold in the
lowest-noise portion of at least one previous image) or likely transient
(above previous detection threshold in the highest-noise portion of at least
one previous image). We find that:

\begin{itemize}

  \item{Likely transients are recovered with a precision of 1.0 and recall of
  0.94;}
  \item{Marginal transients are recovered with a precision of 0.02 and recall 
  of 1.0.}

\end{itemize}

Based on the simplified test described here, we conclude that the algorithm
used to detect likely transients provides a robust way of identifying many
transients with a high resistance to false positives. Further, since all
lightcurve data is retained in the database, the list of marginal transients
provides a key starting point for future manual checking and data mining.

An important limiting factor in this test is the limited resolution at which
the noise maps are stored (i.e.\ just a ``best'' and ``worst'' value for each
image). Recording noise at a more fine-grained level would enable us to
significantly increase the precision with which possible transients are
identified. This is a possible area of future TraP development
(\S\ref{sec:future}).

\subsubsection{Variable sources}

We selected all sources which were initially observed in the first image
(i.e.\ they were not candidates for being marked as new detections) and which
had values of $\eta_\nu$ and $V_\nu$ greater than $0.1$ and $0.01$
respectively.  We constructed an equivalent list based on the simulation
inputs and known image noise level; note that this list excludes some
transient sources which fall below the detection threshold. By combining these
lists, we can calculate the precision and recall
(Eqs.~\ref{eq:precision}~\&~\ref{eq:recall}) as a function of the variability
parameters. These are plotted in Fig.~\ref{fig:FP_properties}.  Above some
limiting value of each threshold, there are no positive detections (either
true or false) so the values of Eqs.~\ref{eq:precision} and \ref{eq:recall}
are undefined; the plots are truncated at this point. Note that for values of
$\eta_\nu \ge 1$ and $V_\nu \ge 0.3$ the precision is 1.0: no false positives
are recorded. Below these values, precision drops rapidly due to noise-based
variation of stable sources.

For all values of $\eta_\nu$ and $V_\nu$ the recall is similarly close to 1.0.
Variations are due to uncertainties introduced by the simulation and
measurement process, which occasionally cause the measured value of the
transient parameters to drop below their predicted values.

\section{Run-time performance}
\label{sec:perf}

As described in sections \ref{sec:lofar:rsm} and \ref{sec:overview}, the TraP
is ultimately intended to perform near real-time analysis of streaming image
data. Although the required rapid imaging capability is not yet available from
LOFAR, we anticipate that other projects---most notably AARTFAAC
\citep{Prasad:2012}---will provide streaming image data in the relatively near
future. It is therefore instructive to consider to what extent the run-time
performance of the TraP is adequate to address such a data stream.

It is worth emphasizing that TraP development to date has focused on
correctness rather than performance. Our aim has been to produce a robust and
well-tested codebase that can then be optimized to address real-time data
processing. We strive to adopt fundamental algorithms which show benign
scaling characteristics to large numbers of images and sources, but emphasize
that the codebase still provides ample opportunity for optimization.

Broadly, we split our consideration of performance characteristics into two
parts, corresponding to the two most computational expensive parts of the
TraP. In \S\ref{sec:perf:sf} we focus on the performance of the Python code,
and particularly that of the computationally intensive source finding
algorithms. In \S\ref{sec:perf:db} we turn to the matter of inserting and
associating measurements in the database and the calculation of per-source
aggregates.

Throughout, we emphasize that there are a large number of tunable parameters
in this analysis, both in terms of the pipeline configuration and the
characteristics of the test data; here, we only give an overview of likely
scaling considerations. For a detailed review of the sourcefinder performance
refer to \citet{Carbone:2015}, and for an in-depth discussion of database
characteristics see \citet{Scheers:2014}.

Throughout this section, the times reported correspond to Python code running
on an Intel Xeon E5-2660v2 CPU with a maximum clock speed of 2.2\,GHz and,
where applicable, interoperating with PostgreSQL 9.3.5 running on an AMD
Opteron 2384 with a maximum clock speed of 2.7\,GHz. We configured PostgreSQL
to make better use of the available system resources by increasing its working
memory (to 100\,MB), its shared buffer (to 2048\,MB) and its checkpoint
interval to 32 segments.

\subsection{Sourcefinder performance}
\label{sec:perf:sf}

\begin{figure}
  \includegraphics[width=\columnwidth]{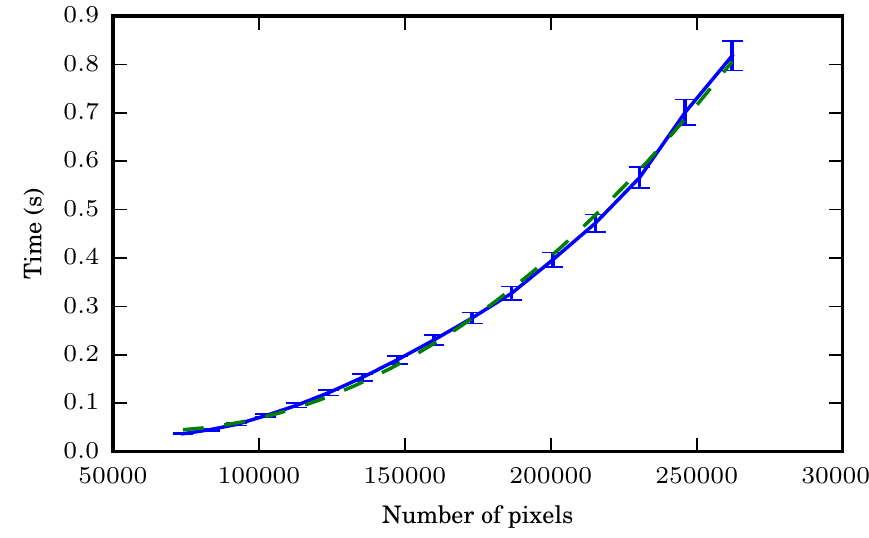}
  \caption{Time taken to calculate the background and RMS maps as a function
  of the number of pixels in the image. The solid line shows the mean measured
  time over 3800 test images; the dashed line, a quadratic fit to the data.}
  \label{fig:sf_map}
\end{figure}

Based on the discussion in \S\ref{sec:components:sf}, we divide the operation
of the sourcefinder into two major components: the calculation of per-image
background and RMS maps, then the identification and measurement of sources
within the image. The former depends on the size of the image, but is
independent of the number of sources within it; the latter increases with
source count.

For each of the 3800 images simulated as described in \S\ref{sec:test:sim} the
time taken to generate background and RMS maps covering the whole image was
measured. The edges of the images were then masked, and timing repeated for
maps covering only the unmasked portion. This process was repeated until only
a small fraction of the image was left unmasked. The times recorded are shown
in Fig.~\ref{fig:sf_map}. For comparison, we also plot the results of a least
squares quadratic fit to the data:
\begin{equation}
t_\mathrm{map} = 1.9\times10^{-11}p^{2} - 2.5\times10^{-6}p + 0.1\,\mathrm{s},
\label{eq:proc:sf:map}
\end{equation}
where $t_\mathrm{map}$ is the time taken to process $p$ pixels. While the
detailed values are system dependent, it is important to note that the
algorithm scales as $\mathcal{O}(N^2)$ in number of pixels.

\begin{figure}
  \includegraphics[width=\columnwidth]{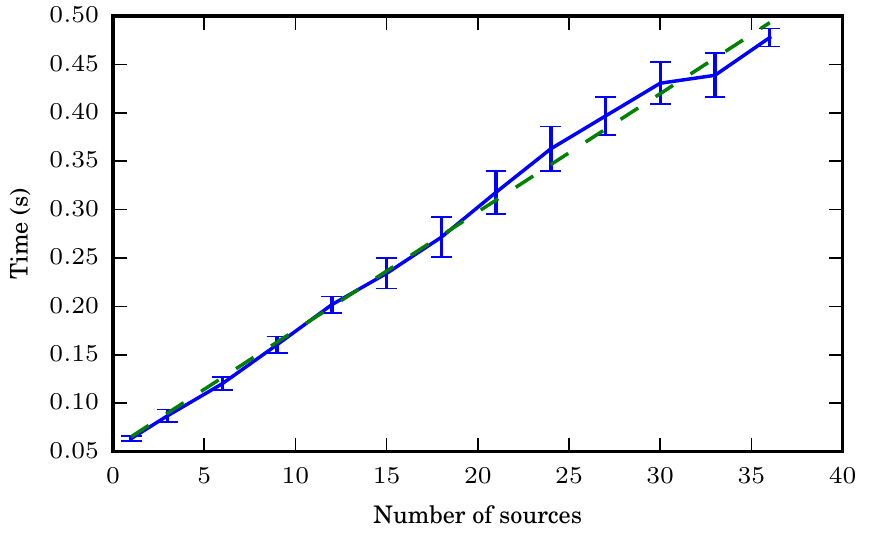}
  \caption{Time taken to find and measure the sources in an image as a
  function of the number of sources found. The solid line shows the mean time
  measured over 3800 test images; the dashed line a linear fit to the data.}
  \label{fig:sf_fit}
\end{figure}

As the unmasked area of each image is decreased, the number of sources which
can be detected and measured within the image also decreases. For each portion
of each image we performed a source finding and measurement step with a
detection threshold of $10\sigma$. The sourcefinder was configured not to
deblend sources (\S\ref{sec:components:sf:deblend}), and to constrain the
shape of the resulting measurements to be equal to the restoring beam
(\S\ref{sec:components:sf:gauss}).

Fig.~\ref{fig:sf_fit} records the total time taken to identify and fit all the
sources in an image as a function of the source number. A linear least squares
fit to the data provides the expression
\begin{equation}
t_\mathrm{fit} = 0.012n + 0.053\,\mathrm{s}
\label{eq:proc:sf:fit}
\end{equation}
for the $t_\mathrm{fit}$ taken to identify and measure $n$ sources. While
again the detailed timings are system-dependent, the key point is the scaling
as $\mathcal{O}(N)$ with number of sources.

\subsection{Database performance}
\label{sec:perf:db}

There are two important axes along which database performance could vary. The
first is with number of images processed: as more data is stored, the number
of source measurements which must be associated and the number of data points
of which aggregates must be calculated increases. For use in a long term
monitoring programme, we require that this accumulation of data does not cause
the database to become slower with time. Secondly, we consider performance as
a function of the number of sources per image: more measurements increase not
only the number of aggregates to be calculated but also the number of
potential source associations.

Artificial source lists representing an artificial sky at arbitrary frequency
and pointing and covering a circular region of radius \degrees{20} were
constructed. Sources were placed on a regular grid within the region.  Each
source was assumed to be a point source, and assigned a random flux density in
the range 1--10\,Jy from a uniform distribution. Sixteen such source lists
were created, containing between 50 and 1200 sources in increments of 50.

For each source population, a set of 100 source measurements was constructed
by perturbing the source position with a Fisher distribution
\citep{Fisher:1953} with concentration parameter $\kappa = 2\times10^9$,
chosen to approximate the systematic position uncertainty of around 5'' which
we have observed to be typical in LOFAR images \citep{Carbone:2015}. This
simulates an observation of the source population.

For each source population, each simulated observation in turn was inserted
and processed (including source association and calculation of aggregate
parameters by the database). The time taken to perform all database operations
was recorded.

\begin{figure}
  \includegraphics[width=\columnwidth]{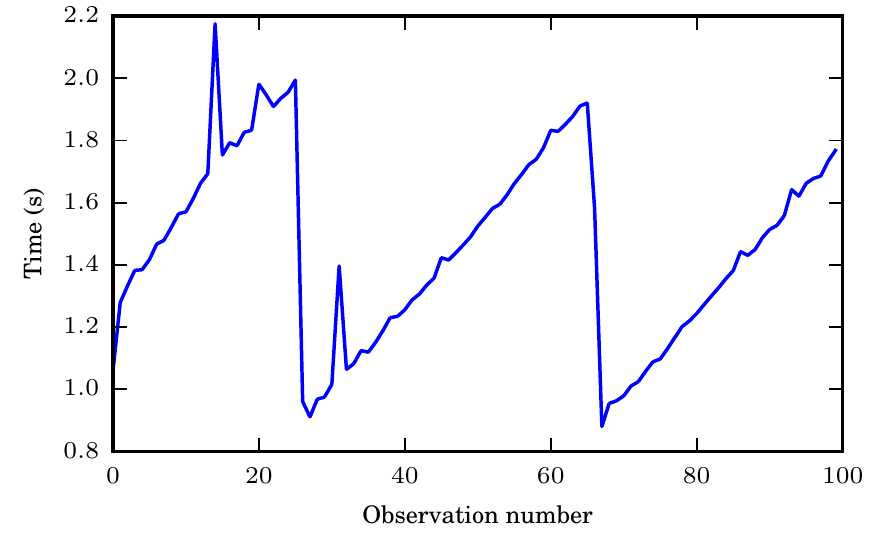}
  \caption{Time taken to process each simulated observation with a population
  of 1100 sources in the database.}
  \label{fig:db_per_image}
\end{figure}

In Fig.~\ref{fig:db_per_image} we show the time taken to perform all the
processing of each simulated observation of 1100 sources as a function of
image number; a similar pattern is observed for all other source counts. The
characteristic ``saw tooth'' pattern in the figure is due to PostgreSQL
periodically checkpointing its write ahead log; other minor variations are
explained by internal housekeeping tasks running within the database and by
varying system and network load over the course of the test. The key result,
though, is that there is no evidence of a systematic increase in processing
time with observation number.

\begin{figure}
  \includegraphics[width=\columnwidth]{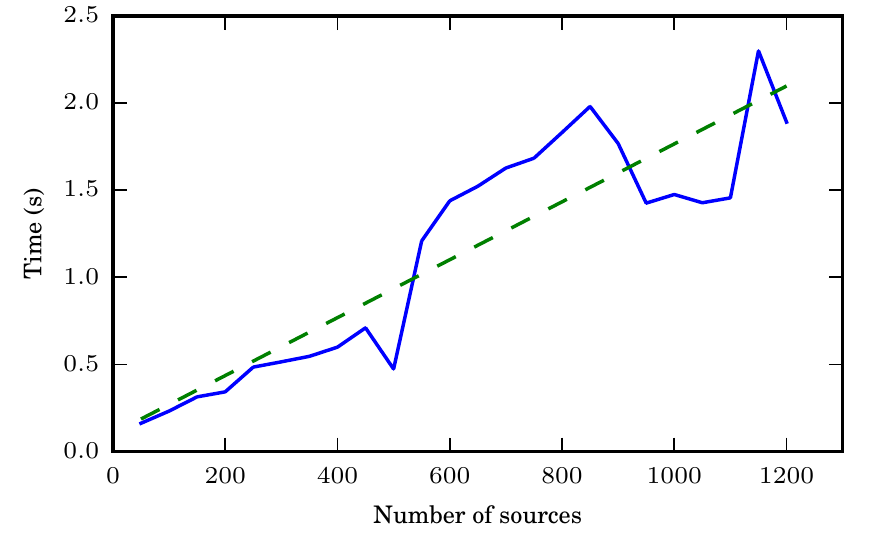}
  \caption{Time taken by the database to process a simulated observation as a
  function of number of sources in the observation. The solid line shows the
  mean value over 100 test observations; the dashed line, a linear function
  for comparison.}
  \label{fig:db_per_nsrc}
\end{figure}

In Fig.~\ref{fig:db_per_nsrc} we show the mean time taken to perform all the
processing of each simulated observation as a function of source count. For
comparison, we also plot the expression
\begin{equation}
t_\mathrm{db} = 0.0017 n + 0.10\,\mathrm{s}
\label{eq:proc:db}
\end{equation}
where $t_\mathrm{db}$ is the time taken to process a simulated observation of
$n$ sources. The detailed timings are, again, system dependent, but it is
important to note the scaling as $\mathcal{O}(N)$ with number of sources per
observation.

\subsection{Practical performance considerations}

\begin{table*}
\begin{center}
\begin{tabular}{lrrrrrrrr}
\toprule
\multicolumn{1}{c}{Configuration} & \multicolumn{1}{c}{Freq.} & \multicolumn{1}{c}{BW}    & \multicolumn{2}{c}{Num.\ stations} & \multicolumn{1}{c}{Max.\ baseline} & \multicolumn{1}{c}{FWHM}   & \multicolumn{1}{c}{Ang.\ res.} & \multicolumn{1}{c}{Pixels per} \\
\multicolumn{1}{c}{name}          & \multicolumn{1}{c}{(MHz)} & \multicolumn{1}{c}{(MHz)} & \multicolumn{1}{c}{Core} & \multicolumn{1}{c}{Remote} & \multicolumn{1}{c}{(km)}          & \multicolumn{1}{c}{(deg.)} & \multicolumn{1}{c}{(asec.)}   & \multicolumn{1}{c}{image}      \\
\midrule
LBA core only       & 60    & 3.6       & 24   & 0  & 3.5 & 9.8   & 294.7    & $1.0\times10^5$ \\
LBA 6\,km baselines & 60    & 3.6       & 24   & 4  & 6.0 & 9.8   & 171.9    & $3.0\times10^5$ \\
LBA NL array        & 60    & 3.6       & 24   & 16 & 121.0         & 9.8   & 8.5      & $1.2\times10^8$ \\
HBA core only       & 150   & 3.6       & 24   & 0  & 3.5           & 3.8   & 117.9    & $9.5\times10^4$ \\
HBA 6\,km baselines & 150   & 3.6       & 24   & 4  & 6.0           & 3.8   & 68.8     & $2.8\times10^5$ \\
HBA NL array        & 150   & 3.6       & 24   & 16 & 121.0         & 3.8   & 3.4      & $1.1\times10^8$ \\
\bottomrule
\end{tabular}
\end{center}
\caption{Parameters of LOFAR observing modes. For each mode, we quote the
angular resolution and full width at half maximum of a single beam and a
bandwidth of 3.6\,MHz, equivalent to 20 subbands.}
\label{tab:config}
\end{table*}

\begin{table*}
\begin{center}
\begin{tabular}{lrrlrrr}
\toprule
\multicolumn{1}{c}{Configuration} & \multicolumn{1}{c}{Integration} & \multicolumn{1}{c}{$5\sigma$ detection limit} & \multicolumn{2}{c}{Source count} & \multicolumn{2}{c}{Processing time (s)} \\
\multicolumn{1}{c}{name}          & \multicolumn{1}{c}{time (s)}    & \multicolumn{1}{c}{(mJy/beam)}            & \multicolumn{1}{c}{Origin} & \multicolumn{1}{c}{Number}  & \multicolumn{1}{c}{Sourcefinder} & \multicolumn{1}{c}{Database} \\
\midrule
LBA core only       &  1 & 2346.5               & VLSS & 27.6 & 0.46 & 0.15 \\
                    & 10 & $\mathcal{C}$ 1066.3 & VLSS & 90.0 & 1.22 & 0.25 \\
\midrule
LBA 6\,km baselines &   1 & 2005.1              & VLSS &  34.9 & 1.58 & 0.16 \\
                    &  10 & 634.1               & VLSS & 196.3 & 3.55 & 0.42 \\
                    & 100 & $\mathcal{C}$ 465.0 & VLSS & 312.6 & 4.98 & 0.62 \\
\midrule
LBA NL array        &     1 & 1395.8  & VLSS &   60.1 & $2.86 \times 10^{5}$ &  0.20 \\
                    &    10 & 441.4   & VLSS &  337.9 & $2.86 \times 10^{5}$ &  0.67 \\
                    &   100 & 139.6   & RSM  &  761.3 & $2.86 \times 10^{5}$ &  1.37 \\
                    &  1000 & 44.2    & WENSS & 2043.6 & $2.86 \times 10^{5}$ & 3.50 \\
                    & 10000 & 14.0    & FIRST & 3537.0 & $2.86 \times 10^{5}$ & 5.99 \\
\midrule
HBA core only       &  1 & $\mathcal{C}$ 137.0 & LOFAR & 45.3 & 0.67 & 0.18 \\
\midrule
HBA 6\,km baselines &  1 & 102.0              & LOFAR & 70.5 & 1.86 & 0.22 \\
                    & 10 & $\mathcal{C}$ 59.7 & LOFAR & 157.4 & 2.92 & 0.37 \\
\midrule
HBA NL array        &     1 & 82.8 & LOFAR &    96.4 & $2.53 \times 10^5$ &  0.26 \\
                    &    10 & 26.2 & WENSS &   323.0 & $2.53 \times 10^5$ &  0.64 \\
                    &   100 &  8.3 & FIRST &   449.5 & $2.53 \times 10^5$ &  0.85 \\
                    &  1000 &  2.6 & FIRST &  2516.8 & $2.53 \times 10^5$ &  4.29 \\
                    & 10000 &  0.8 & FIRST & 14153.5 & $2.53 \times 10^5$ & 23.64 \\
\bottomrule
\end{tabular}
\end{center}
\caption{Source counts and processing times predicted for a single image using
each of the LOFAR configurations described in Tab.~\ref{tab:config} at a range
of integration times. The symbol $\mathcal{C}$ indicates that the detection
limit was set by confusion noise rather than image sensitivity.}
\label{tab:proctime}
\end{table*}

We conclude this discussion of pipeline runtime performance by comparing the
measured TraP performance to potential LOFAR transient monitoring strategies.

The initial Radio Sky Monitor strategy \citep{Fender:2014} is to use six beams
from LOFAR to tile out a wide area on the sky. Each beam consists of four
frequency bands, which are imaged separately. Assuming LOFAR is operating in 8
bit mode (\S\ref{sec:lofar:ilt}), each band contains 20 subbands and provides a
bandwidth of 3.6\,MHz.

In Tab.~\ref{tab:config} we list the parameters of each of six major LOFAR
observing modes and provide the full width at half maximum and the angular
resolution of a single image constructed using the survey strategy described.
The observing modes include using only the core LOFAR stations, using the full
Dutch LOFAR array, and using only that subset of the full array which contains
baselines no more than 6\,km in length. This latter configuration has been
shown to provide a good compromise between image fidelity and processing time
in early LOFAR observations and has been used for initial RSM observations.

In Tab.~\ref{tab:proctime} we provide estimates of the $5\sigma$ detection
limit and corresponding source count for each of the configurations at a range
of integration times. Sensitivity and confusion limits were estimated using
the online LOFAR Image Noise Calculator\footnote{Version 0.31;
\url{http://www.astron.nl/~heald/test/sens2.php}}. Following current standard
LOFAR observing practice, when using the high band remote stations were
``tapered'', reducing sensitivity while increasing field of view to match that
of the stations in the core. Where possible, source counts were taken from
preliminary analysis of LOFAR RSM data \citep{Fender:2014}; otherwise, they
were estimated by extrapolating from other surveys assuming a spectral index
of $-0.7$: for each configuration, the survey best approximating the spatial
resolution was selected from FIRST \citep[Faint Images of the Radio Sky at
Twenty-cm;][]{Becker:1994}, NVSS
\citep{Condon:1998}, VLSS \citep{Cohen:2007} and WENSS \citep{Rengelink:1997}.

Also in Tab.~\ref{tab:proctime} we estimate the processing time for each image
through both the sourcefinder ($t_{\mathrm{map}} + t_{\mathrm{fit}}$) and the
database ($t_{\mathrm{db}}$). Note that these equations reflect the particular
characteristics of the systems used for benchmarking and can provide only
guideline performance estimates. Further note that this estimate includes only
the major pipeline components identified above; it is reasonable to expect
modest overheads from other parts of the pipeline.

Note that for most image types excluding the full array the processing time per
image is considerably less than the integration time for that image. This
indicates that processing a real-time stream of images of this type is tractable
on the systems used for benchmarking.

As described above, the likely operation configuration consists not of a single
image stream but of four bands in each of six beams: a factor of 24 greater than
the figures quoted. For all 1\,s, and some longer, integrations, this increases
the total processing time to greater than the integration time. However, three
factors mitigate this:

\begin{itemize}

  \item{As described in \S\ref{sec:implement:parallel}, we support parallel
  (and, optionally, distributed) sourcefinder operation: by distributing the
  processing time required over multiple CPUs, we achieve a near-linear
  reduction in wall-clock time.}

  \item{By starting to search for sources in timestep $t_{n+1}$ before
  timestep $t_{n}$ has finished processing, we can ensure that both
  sourcefinding and database systems are efficiently occupied, rather than
  synchronously waiting for each other.}

  \item{The hardware platform used for benchmarking the database is several
  years old; more powerful systems are now available which will provide a
  significant constant factor improvement in database throughput.}

\end{itemize}

We also emphasize that the real-time LOFAR imaging mode is still under
development (\S\ref{sec:lofar:rsm}): all current processing is carried out
off-line. The run-time performance of the TraP is significantly better than that
of the imaging tools used to provide it with input in this mode.

Finally, we reiterate that TraP development to date has focused on correctness
over performance. There is much still to be gained in terms of optimization of
individual algorithms, efficiency of implementation (e.g.\ replacing core
Python loops with Cython equivalents) and best exploiting the performance of
high level tools (e.g.\ the potential gains of the MonetDB system over
PostgreSQL as described in \S\ref{sec:implement:db:systems}).

\section{Releases and future development}
\label{sec:future}

This manuscript describes release 2.0 of the TraP, dating from December 2014.
With this release, the TraP became an open source project, and therefore
freely available to download from our GitHub
repository\footnote{\url{https://github.com/transientskp/tkp/}} under a
two-clause
BSD-style\footnote{\url{http://opensource.org/licenses/BSD-2-Clause}} license.
If you use the TraP, or code derived from it, in a paper or other publication,
we request that you cite this work.

After this release, development will continue. Future releases of the TraP are
expected to expand upon the current functionality to offer features such as:

\begin{itemize}

  \item{Performance optimization in support of real-time streaming transient monitoring;}
  \item{Automatic preliminary classification of detected transients;}
  \item{Support for multi-terabyte lightcurve archives and enhanced archive data-mining and visualization;}
  \item{Automatic cross-correlation of TraP detected sources with known catalogue sources across a range of wavelengths;}
  \item{Improved methods for apportioning flux density from a single measurement among multiple sources on association;}
  \item{Full Stokes support, for both identifying and classifying sources;}
  \item{Higher resolution and more flexible noise maps;}
  \item{Construction and usage of a deep average image of the surveyed area.}

\end{itemize}

\section{Conclusions}

The current and next generation of astronomical survey facilities, across a
wide range of wavelengths but particularly in the radio, provide an
opportunity to explore the transient and variable sky in powerful and
unprecedented ways. This is especially true of LOFAR, which combines a
remarkably wide field of view with unique sensitivity to low radio frequencies
and a flexible, software-driven architecture.  However, identifying transients
in the massive data volumes produced by these instruments is challenging.

This manuscript has described our attempt to rise to this challenge in the
form of the LOFAR Transients Pipeline. It combines a flexible,
high-performance architecture with robust analysis tools in a well tested and
documented package. We have shown how it can both be used to generate alert
messages as new transients are discovered and to populate a database of
lightcurves of potential transients for offline analysis. We have demonstrated
that it is capable of accurately recovering a known population of transients
from simulated LOFAR observations.

The TraP is now being used in support of ongoing LOFAR observing campaigns.
However, development continues, and we are actively expanding its
capabilities, both to better address ongoing LOFAR operations and to increase
its applicability to other instruments and wavelength regimes.  The codebase
is open source and freely available; we actively invite you to join us in
improving it.

\section{Acknowledgements}

This work has made use of a variety of Python libraries, including Matplotlib
\citep{Hunter:2007}. The authors acknowledge support from the European
Research Council via Advanced Investigator Grants 247295 and 267697. BSch is
in part funded by the research programme of the Netherlands eScience
Center\footnote{\url{http://www.nlesc.nl/}}. CS and HG acknowledge financial
support from the UnivEarthS Labex program of Sorbonne Paris Cit\'{e}
(ANR-10-LABX-0023 and ANR-11-IDEX-0005-02). JWTH acknowledges funding from an
NWO Vidi fellowship and ERC Starting Grant ``DRAGNET'' (337062).

\section*{References}

\bibliographystyle{elsarticle-harv}
\bibliography{trap}

\begin{thebibliography}{112}
\expandafter\ifx\csname natexlab\endcsname\relax\def\natexlab#1{#1}\fi
\expandafter\ifx\csname url\endcsname\relax
  \def\url#1{\texttt{#1}}\fi
\expandafter\ifx\csname urlprefix\endcsname\relax\def\urlprefix{URL }\fi

\bibitem[{{Alard} and {Lupton}(1998)}]{Alard:1998}
{Alard}, C., {Lupton}, R.~H., Aug. 1998. {A Method for Optimal Image
  Subtraction}. Astrophys. J. 503, 325--331.

\bibitem[{{Allan} and {Denny}(2009)}]{Allan:2009}
{Allan}, A., {Denny}, R.~B., Aug. 2009. {VOEvent Transport Protocol, Version
  1.1}. {Note}, {International Virtual Observatory Alliance}.

\bibitem[{Amdahl(1967)}]{Amdahl:1967}
Amdahl, G.~M., 1967. Validity of the single processor approach to achieving
  large scale computing capabilities. In: Proceedings of the April 18-20, 1967,
  Spring Joint Computer Conference. AFIPS '67 (Spring). ACM, New York, NY, USA,
  pp. 483--485.

\bibitem[{{Anderson} et~al.(2014)}]{Anderson:2014}
{Anderson}, G.~E., et~al., 2014. {Probing the bright radio flare and afterglow
  of GRB 130427A with the Arcminute Microkelvin Imager}. Mon. Not. R. Astron.
  Soc. 440, 2059--2065.

\bibitem[{{Astropy Collaboration} et~al.(2013)}]{Astropy:2013}
{Astropy Collaboration}, et~al., Oct. 2013. {Astropy: A community Python
  package for astronomy}. Astron. Astrophys. 558, A33.

\bibitem[{{Atwood} et~al.(2009)}]{Atwood:2009}
{Atwood}, W.~B., et~al., Jun. 2009. {The Large Area Telescope on the Fermi
  Gamma-Ray Space Telescope Mission}. Astrophys. J. 697, 1071--01102.

\bibitem[{{Bannister} et~al.(2011)}]{Bannister:2011}
{Bannister}, K.~W., et~al., Mar. 2011. {A 22-yr southern sky survey for
  transient and variable radio sources using the Molonglo Observatory Synthesis
  Telescope}. Mon. Not. R. Astron. Soc. 412, 634--664.

\bibitem[{{Banyer} et~al.(2012){Banyer}, {Murphy}, and {the VAST
  Collaboration}}]{Banyer:2012}
{Banyer}, J., {Murphy}, T., {the VAST Collaboration}, 2012. {VAST -- a
  real-time pipeline for detecting radio transients and variables on the
  Australian SKA Pathfinder (ASKAP) telescope}. In: {Ballester}, P. (Ed.),
  Astronomical Data Analysis Software and Systems XXI. Vol. 461 of Astronomical
  Society of the Pacific Conference Series. p. 725.

\bibitem[{{Becker} et~al.(1994){Becker}, {White}, and {Helfand}}]{Becker:1994}
{Becker}, R.~H., {White}, R.~L., {Helfand}, D.~J., 1994. {The VLA's FIRST
  Survey}. In: {Crabtree}, D.~R., {Hanisch}, R.~J., {Barnes}, J. (Eds.),
  Astronomical Data Analysis Software and Systems III. Vol.~61 of Astronomical
  Society of the Pacific Conference Series. p. 165.

\bibitem[{Bell et~al.(2011)}]{Bell:2011}
Bell, M.~E., et~al., 2011. {An automated archival Very Large Array transients
  survey}. Mon. Not. R. Astron. Soc. 415, 2--10.

\bibitem[{{Bell} et~al.(2014)}]{Bell:2014}
{Bell}, M.~E., et~al., Feb. 2014. {A survey for transients and variables with
  the Murchison Widefield Array 32-tile prototype at 154 MHz}. Mon. Not. R.
  Astron. Soc. 438, 352--367.

\bibitem[{{Benjamini} and {Hochberg}(1995)}]{Benjamini:1995}
{Benjamini}, Y., {Hochberg}, Y., 1995. {Controlling the false discovery rate: a
  practical and powerful approach to multiple testing}. {J. Roy. Statist. Soc.
  Ser. B} 57, 289--300.

\bibitem[{{Bertin} and {Arnouts}(1996)}]{Bertin:1996}
{Bertin}, E., {Arnouts}, S., Jun. 1996. {SExtractor: Software for source
  extraction.} Astron. Astrophys. Suppl. Ser. 117, 393--404.

\bibitem[{{Bolch}(1968)}]{Bolch:1968}
{Bolch}, B.~W., 1968. {More on unbiased estimation of the standard deviation}.
  {The American Statistician} 22, 27.

\bibitem[{{Booth} and {Jonas}(2012)}]{Booth:2012}
{Booth}, R.~S., {Jonas}, J.~L., Mar. 2012. {An Overview of the MeerKAT
  Project}. African Skies 16, 101.

\bibitem[{{Bower} et~al.(2011)}]{Bower:2011}
{Bower}, G.~C., et~al., Oct. 2011. {The Allen Telescope Array Pi GHz Sky Survey
  II. Daily and Monthly Monitoring for Transients and Variability in the
  Bo{\"o}tes Field}. Astrophys. J. 739, 76.

\bibitem[{Broderick et~al.(in prep.)}]{Broderick:2014}
Broderick, J., et~al., in prep.

\bibitem[{{Brown} et~al.(2013)}]{Brown:2013}
{Brown}, T.~M., et~al., Sep. 2013. {Las Cumbres Observatory Global Telescope
  Network}. Publ. Astron. Soc. Pac. 125, 1031--1055.

\bibitem[{{Carbone} et~al.(in prep.)}]{Carbone:2015}
{Carbone}, D., et~al., in prep. {PySE: Software for Extracting Sources from
  LOFAR Radio Telescope Images}.

\bibitem[{{Carbone} et~al.(submitted)}]{Carbone:2015a}
{Carbone}, D., et~al., submitted. {New methods to constrain the radio transient
  rate: results from a survey of four fields with LOFAR}.

\bibitem[{{Codd}(1970)}]{Codd:1970}
{Codd}, E.~F., Jun. 1970. {A relational model of data for large shared data
  banks}. Commun. ACM 13, 377--387.

\bibitem[{{Coenen} et~al.(2014)}]{Coenen:2014}
{Coenen}, T., et~al., Oct. 2014. {The LOFAR pilot surveys for pulsars and fast
  radio transients}. Astron. Astrophys. 570, A60.

\bibitem[{{Cohen} et~al.(2007)}]{Cohen:2007}
{Cohen}, A.~S., et~al., Sep. 2007. {The VLA Low-Frequency Sky Survey}. Astron.
  J. 134, 1245--1262.

\bibitem[{{Condon}(1997)}]{Condon:1997}
{Condon}, J.~J., 1997. Errors in elliptical {G}aussian fits. Publ. Astron. Soc.
  Pac. 109, 166--172.

\bibitem[{{Condon} et~al.(1998)}]{Condon:1998}
{Condon}, J.~J., et~al., May 1998. {The NRAO VLA Sky Survey}. Astron. J. 115,
  1693--1716.

\bibitem[{{Croft} et~al.(2011)}]{Croft:2011}
{Croft}, S., et~al., Apr. 2011. {The Allen Telescope Array Twenty-centimeter
  Survey---A 700-square-degree, Multi-epoch Radio Data Set. II. Individual
  Epoch Transient Statistics}. Astrophys. J. 731, 34.

\bibitem[{{de Bruyn} et~al.(2009){de Bruyn}, {Bernardi}, and {Lofar
  Eor-Team}}]{deBruyn:2009}
{de Bruyn}, A.~G., {Bernardi}, G., {Lofar Eor-Team}, Sep. 2009. {The First Deep
  WSRT 150\,MHz Full Polarization Observations}. In: {Saikia}, D.~J., {Green},
  D.~A., {Gupta}, Y., {Venturi}, T. (Eds.), The Low-Frequency Radio Universe.
  Vol. 407 of Astron. Soc. Pac. Conf. Ser. p.~3.

\bibitem[{{de Ruiter} et~al.(1977){de Ruiter}, {Willis}, and
  {Arp}}]{deRuiter:1977}
{de Ruiter}, H.~R., {Willis}, A.~G., {Arp}, H.~C., May 1977. {A Westerbork
  1415\,MHz survey of background radio sources. II - Optical identifications
  with deep IIIA-J plates}. Astron. Astrophys. Suppl. Ser. 28, 211--293.

\bibitem[{{Denneau} et~al.(2013)}]{Denneau:2013}
{Denneau}, L., et~al., Apr. 2013. {The Pan-STARRS Moving Object Processing
  System}. Publ. Astron. Soc. Pac. 125, 357--395.

\bibitem[{{Dent}(1965)}]{Dent:1965}
{Dent}, W.~A., Jun. 1965. {Quasi-Stellar Sources: Variation in the Radio
  Emission of 3C 273}. Science 148, 1458--1460.

\bibitem[{Dewdney et~al.(2010)}]{Dewdney:2010}
Dewdney, P., et~al., 2010. {SKA Memo 130: SKA Phase 1: Preliminary System
  Description}.

\bibitem[{{Drake} et~al.(2009)}]{Drake:2009}
{Drake}, A.~J., et~al., May 2009. {First Results from the Catalina Real-Time
  Transient Survey}. Astrophys. J. 696, 870--884.

\bibitem[{Fender et~al.(2006)}]{Fender:2006}
Fender, R., et~al., 2006. {The LOFAR Transients Key Project}. In: {Proceedings
  of the VI Microquasar Workshop: Microquasars and Beyond}. PoS(MQW6)014.

\bibitem[{{Fender} et~al.(in~prep.)}]{Fender:2014}
{Fender}, R., et~al., in~prep. {The LOFAR Radio Sky Monitor}.

\bibitem[{{Fender} and {Bell}(2011)}]{Fender:2011}
{Fender}, R.~P., {Bell}, M.~E., Sep. 2011. {Radio transients: an antediluvian
  review}. Bulletin of the Astronomical Society of India 39, 315--332.

\bibitem[{{Fisher}(1953)}]{Fisher:1953}
{Fisher}, R., 1953. {Dispersion on a sphere}. {Proc. Roy. Soc. London Ser. A.}
  219, 295--305.

\bibitem[{{Fishman}(1992)}]{Fishman:1992}
{Fishman}, G.~J., 1992. {BATSE -- The burst and transient source experiment on
  the Gamma Ray Observatory}. pp. 265--272.

\bibitem[{{Frail} et~al.(1999){Frail}, {Kulkarni}, and {Bloom}}]{Frail:1999}
{Frail}, D.~A., {Kulkarni}, S.~R., {Bloom}, J.~S., Mar. 1999. {An outburst of
  relativistic particles from the soft {$\gamma$}-ray repeater SGR1900+14}.
  Nature 398, 127--129.

\bibitem[{{Frail} et~al.(1997)}]{Frail:1997}
{Frail}, D.~A., et~al., Sep. 1997. {The radio afterglow from the {$\gamma$}-ray
  burst of 8 May 1997}. Nature 389, 261--263.

\bibitem[{{Frail} et~al.(2012)}]{Frail:2012}
{Frail}, D.~A., et~al., Mar. 2012. {A revised view of the transient radio sky}.
  Astrophys. J. 747, 70.

\bibitem[{Freund and Williams(2010)}]{Freund:2010}
Freund, J.~E., Williams, F.~J., 2010. Outline of Basic Statistics: Dictionary
  and Formulas. Courier Corporation.

\bibitem[{{Gaensler} and {Hunstead}(2000)}]{Gaensler:2000}
{Gaensler}, B.~M., {Hunstead}, R.~W., Apr. 2000. {Long-term Monitoring of
  Molonglo Calibrators}. Publ. Astron. Soc. Aus. 17, 72--82.

\bibitem[{{Gautschi}(1972)}]{Gautschi:1972}
{Gautschi}, W., 1972. {Error function and Fresnel integrals}. Dover.

\bibitem[{{Gehrels} et~al.(2004)}]{Gehrels:2004}
{Gehrels}, N., et~al., Aug. 2004. {The Swift Gamma-Ray Burst Mission}.
  Astrophys. J. 611, 1005--1020.

\bibitem[{{Graham} et~al.(2014)}]{Graham:2014}
{Graham}, M., et~al., May 2014. {Employing SimpleTimeSeries for representing
  time series, version 1.0}. {Note}, {International Virtual Observatory
  Alliance}.

\bibitem[{Gray et~al.(2006)Gray, Nieto-Santisteban, and Szalay}]{Gray:2006}
Gray, J., Nieto-Santisteban, M.~A., Szalay, A.~S., 2006. {The Zones Algorithm
  for Finding Points-Near-a-Point or Cross-Matching Spatial Datasets}. MSR TR
  2006 52, {Microsoft and The Johns Hopkins University}.

\bibitem[{{Gregory} et~al.(1972)}]{Gregory:1972}
{Gregory}, P.~C., et~al., Oct. 1972. {The nature of the first Cygnus X-3 radio
  outburst}. Nature Physical Science 239, 114--117.

\bibitem[{{Greisen}(2003)}]{Greisen:2003}
{Greisen}, E.~W., Mar. 2003. {AIPS, the VLA, and the VLBA}. Information
  Handling in Astronomy - Historical Vistas 285, 109.

\bibitem[{{Hancock} et~al.(2012)}]{Hancock:2012}
{Hancock}, P.~J., et~al., May 2012. {Compact continuum source finding for next
  generation radio surveys}. Mon. Not. R. Astron. Soc. 422, 1812--1824.

\bibitem[{{Harmon} et~al.(2002)}]{Harmon:2002}
{Harmon}, B.~A., et~al., Jan. 2002. {The Burst and Transient Source Experiment
  Earth Occultation Technique}. Astrophys. J., Suppl. Ser. 138, 149--183.

\bibitem[{{Heald} et~al.(2010)}]{Heald:2010}
{Heald}, G., et~al., Aug. 2010. {Progress with the LOFAR Imaging Pipeline}. In:
  {Proceedings of the ISKAF2010 Science Meeting}. PoS(ISKAF2010)057.

\bibitem[{{Heald} et~al.(2011)}]{Heald:2011}
{Heald}, G., et~al., Dec. 2011. {LOFAR: Recent Imaging Results and Future
  Prospects}. Journal of Astrophysics and Astronomy 32, 589--598.

\bibitem[{{Heald} et~al.(2014)}]{Heald:2014}
{Heald}, G., et~al., 2014. {The LOFAR Multifrequency Snapshot Sky Survey
  (MSSS): 1. Survey description and first results}. Astron.
  Astrophys.Submitted.

\bibitem[{Hendricks and Robey(1936)}]{Hendricks:1936}
Hendricks, W.~A., Robey, K.~W., 1936. The sampling distribution of the
  coefficient of variation. The Annals of Mathematical Statistics 7~(3),
  129--132.

\bibitem[{{H\"ogbom}(1974)}]{Hogbom:1974}
{H\"ogbom}, J.~A., 1974. {Aperture Synthesis with a non-regular distribution of
  interferometer baselines}. Astron. Astrophys. Suppl. Ser. 15, 417--426.

\bibitem[{{Hopkins} et~al.(2002)}]{Hopkins:2002}
{Hopkins}, A.~M., et~al., Feb. 2002. {A New Source Detection Algorithm Using
  the False-Discovery Rate}. Astron. J. 123, 1086--1094.

\bibitem[{{Hopkins} et~al.(2003)}]{Hopkins:2003}
{Hopkins}, A.~M., et~al., Feb. 2003. {The Phoenix Deep Survey: The 1.4 GHz
  Microjansky Catalog}. Astron. J. 125, 465--477.

\bibitem[{{Hopkins} et~al.(in prep.)}]{Hopkins:2015}
{Hopkins}, A.~M., et~al., in prep. {The ASKAP/EMU Source Finding Data
  Challenge}.

\bibitem[{Hunter(2007)}]{Hunter:2007}
Hunter, J.~D., 2007. Matplotlib: A 2d graphics environment. Computing In
  Science \& Engineering 9~(3), 90--95.

\bibitem[{Idreos et~al.(2012)}]{Idreos:2012}
Idreos, S., et~al., Mar. 2012. M{onetDB}: {Two} {Decades} {Of} {Research} {In}
  {Column-}{Oriented} {Database} {Architectures}. IEEE Data Engineering
  Bulletin 35~(1), 40--45.

\bibitem[{{Ivezi\'{c}} et~al.(2014)}]{Ivezic:2014}
{Ivezi\'{c}}, {\v Z}., et~al., 2014. {LSST: From science drivers to reference
  design and anticipated data products, version 3.1}. arXiv:0805.2366.

\bibitem[{Jenet and Gil(2003)}]{Jenet:2003}
Jenet, F.~A., Gil, J., 2003. Using the intensity modulation index to test
  pulsar radio emission models. The Astrophysical Journal Letters 596~(2),
  L215.

\bibitem[{Kent(1983)}]{Kent:1983}
Kent, W., Feb. 1983. A simple guide to five normal forms in relational database
  theory. Commun. ACM 26~(2), 120--125.

\bibitem[{{Kesteven} et~al.(1977){Kesteven}, {Bridle}, and
  {Brandie}}]{Kesteven:1977}
{Kesteven}, M.~J.~L., {Bridle}, A.~H., {Brandie}, G.~W., Aug. 1977.
  {Variability of extragalactic sources at 2.7 GHz. III - The nature of the
  variations in different source classes}. Astron. J. 82, 541--556.

\bibitem[{{Koposov} and {Bartunov}(2006)}]{Koposov:2006}
{Koposov}, S., {Bartunov}, O., Jul. 2006. {Q3C, Quad Tree Cube -- The new
  sky-indexing concept for huge astronomical catalogues and its realization for
  main astronomical queries (cone search and xmatch) in Open Source Database
  PostgreSQL}. In: {Gabriel}, C., {Arviset}, C., {Ponz}, D., {Enrique}, S.
  (Eds.), Astronomical Data Analysis Software and Systems XV. Vol. 351 of
  Astronomical Society of the Pacific Conference Series. p. 735.

\bibitem[{Lande(1977)}]{Lande:1977}
Lande, R., 1977. On comparing coefficients of variation. Systematic Biology
  26~(2), 214--217.

\bibitem[{{Law} et~al.(2011)}]{Law:2011}
{Law}, C.~J., et~al., Nov. 2011. {Millisecond Imaging of Radio Transients with
  the Pocket Correlator}. Astrophys. J. 742, 12.

\bibitem[{{Law} et~al.(2009)}]{Law:2009}
{Law}, N.~M., et~al., Dec. 2009. {The Palomar Transient Factory: System
  Overview, Performance, and First Results}. Publ. Astron. Soc. Pac. 121,
  1395--1408.

\bibitem[{{Levan} et~al.(2011)}]{Levan:2011}
{Levan}, A.~J., et~al., Jul. 2011. {An extremely luminous panchromatic outburst
  from the nucleus of a distant galaxy}. Science 333, 199--202.

\bibitem[{{Levine} et~al.(1996)}]{Levine:1996}
{Levine}, A.~M., et~al., Sep. 1996. {First Results from the All-Sky Monitor on
  the Rossi X-Ray Timing Explorer}. Astrophys. J. 469, L33.

\bibitem[{{Loose}(2008)}]{Loose:2008}
{Loose}, G.~M., Aug. 2008. {LOFAR Self-Calibration Using a Blackboard Software
  Architecture}. In: {Argyle}, R.~W., {Bunclark}, P.~S., {Lewis}, J.~R. (Eds.),
  Astronomical Data Analysis Software and Systems XVII. Vol. 394 of
  Astronomical Society of the Pacific Conference Series. p.~91.

\bibitem[{{Lyubarsky}(2014)}]{Lyubarsky:2014}
{Lyubarsky}, Y., Jul. 2014. {A model for fast extragalactic radio bursts}. Mon.
  Not. R. Astron. Soc. 442, L9--L13.

\bibitem[{{McDowell} et~al.(2015)}]{McDowell:2015}
{McDowell}, J., et~al., Feb. 2015. {IVOA Spectral Data Model, Version 2.0}.
  {Proposed Recommendation}, {International Virtual Observatory Alliance}.

\bibitem[{McKay(1932)}]{McKay:1932}
McKay, A.~T., 1932. Distribution of the coefficient of variation and the
  extended "t" distribution. Journal of the Royal Statistical Society 95~(4),
  695--698.

\bibitem[{{Meegan} et~al.(2009)}]{Meegan:2009}
{Meegan}, C., et~al., Sep. 2009. {The Fermi Gamma-ray Burst Monitor}.
  Astrophys. J. 702, 791--804.

\bibitem[{{Mooley} et~al.(2013)}]{Mooley:2013}
{Mooley}, K.~P., et~al., May 2013. {Sensitive Search for Radio Variables and
  Transients in the Extended Chandra Deep Field South}. Astrophys. J. 768, 165.

\bibitem[{{Mor\'e}(1977)}]{More:1977}
{Mor\'e}, J.~J., 1977. The {L}evenberg-{M}arquardt algorithm: Implementation
  and theory. In: {Watson}, G.~A. (Ed.), {Numerical Analysis}. Vol. 630 of
  {Lecture Notes in Mathematics}. pp. 105--116.

\bibitem[{{Murphy} et~al.(2013)}]{Murphy:2013}
{Murphy}, T., et~al., Feb. 2013. {VAST: An ASKAP Survey for Variables and Slow
  Transients}. Publ. Astron. Soc. Aus. 30, 6.

\bibitem[{{Narayan}(1992)}]{Narayan:1992}
{Narayan}, R., Oct. 1992. {The Physics of Pulsar Scintillation}. Royal Society
  of London Philosophical Transactions Series A 341, 151--165.

\bibitem[{{Nijboer} et~al.(2009){Nijboer}, {Pandey-Pommier}, and {de
  Bruyn}}]{Nijboer:2009}
{Nijboer}, R.~J., {Pandey-Pommier}, M., {de Bruyn}, A.~G., 2009. {LOFAR imaging
  capabilities and system sensitivity}. Memo 113, Square Kilometre Array
  Project.

\bibitem[{{O'Mullane} et~al.(2005)}]{OMullane:2005}
{O'Mullane}, W., et~al., Feb. 2005. {Batch is back: CasJobs, serving multi-TB
  data on the Web}. eprint arXiv:cs/0502072.

\bibitem[{P\'erez and Granger(2007)}]{Perez:2007}
P\'erez, F., Granger, B.~E., May 2007. {IP}ython: a system for interactive
  scientific computing. Computing in Science and Engineering 9~(3), 21--29.

\bibitem[{{Prasad} and {Wijnholds}(2012)}]{Prasad:2012}
{Prasad}, P., {Wijnholds}, S.~J., May 2012. {AARTFAAC: Towards a 24x7, All-sky
  Monitor for LOFAR}. In: {Proceedings of `New Windows on Transients across the
  Universe': A discussion meeting of the Royal Society, London}.

\bibitem[{{Rau} et~al.(2009)}]{Rau:2009}
{Rau}, A., et~al., Dec. 2009. {Exploring the Optical Transient Sky with the
  Palomar Transient Factory}. Publ. Astron. Soc. Pac. 121, 1334--1351.

\bibitem[{{Rengelink} et~al.(1997)}]{Rengelink:1997}
{Rengelink}, R.~B., et~al., Aug. 1997. {The Westerbork Northern Sky Survey
  (WENSS). I. A 570 square degree Mini-Survey around the North Ecliptic Pole}.
  Astron. Astrophys. Suppl. Ser. 124, 259--280.

\bibitem[{{Rickett}(1990)}]{Rickett:1990}
{Rickett}, B.~J., 1990. {Radio propagation through the turbulent interstellar
  plasma}. Ann. Rev. Astron. Astrophys. 28, 561--605.

\bibitem[{{Rowlinson} et~al.(in prep.)}]{Rowlinson:2015}
{Rowlinson}, A.~R., et~al., in prep. {Methodology for optimising the LOFAR
  Transients Pipeline}. Astron. Comp.

\bibitem[{{Rutledge} et~al.(2000){Rutledge}, {Brunner}, {Prince}, and
  {Lonsdale}}]{Rutledge:2000}
{Rutledge}, R.~E., {Brunner}, R.~J., {Prince}, T.~A., {Lonsdale}, C., Nov.
  2000. {XID: Cross-Association of ROSAT/Bright Source Catalog X-Ray Sources
  with USNO A-2 Optical Point Sources}. Astrophys. J., Suppl. Ser. 131,
  335--353.

\bibitem[{{Sagiv} and {Waxman}(2002)}]{Sagiv:2002}
{Sagiv}, A., {Waxman}, E., Aug. 2002. {Collective Processes in Relativistic
  Plasma and Their Implications for Gamma-Ray Burst Afterglows}. Astrophys. J.
  574, 861--872.

\bibitem[{{Scheers}(2011)}]{Scheers:2011}
{Scheers}, L.~H.~A., 2011. {Transient and Variable Radio Sources in the LOFAR
  Sky}. Ph.D. thesis, {University of Amsterdam}.
\newline\urlprefix\url{http://hdl.handle.net/11245/1.328956}

\bibitem[{{Scheers} et~al.(in prep.)}]{Scheers:2014}
{Scheers}, L.~H.~A., et~al., in prep. {LOFAR Transients Key Project Archival
  Database}.

\bibitem[{{Seaman} et~al.(2011)}]{Seaman:2011}
{Seaman}, R., et~al., Jul. 2011. Sky event reporting metadata, version 2.0.
  {Recommendation}, {International Virtual Observatory Alliance}.

\bibitem[{{Spreeuw}(2010)}]{Spreeuw:2010}
{Spreeuw}, J.~N., 2010. {Search and detection of low frequency radio
  transients}. Ph.D. thesis, {University of Amsterdam}.
\newline\urlprefix\url{http://hdl.handle.net/11245/1.324881}

\bibitem[{Staley et~al.(2013)}]{Staley:2013}
Staley, T.~D., et~al., 2013. {Automated rapid follow-up of Swift gamma-ray
  burst alerts at 15\,GHz with the AMI Large Array}. Mon. Not. R. Astron. Soc.
  428, 3114--3120.

\bibitem[{{Stappers} et~al.(2011)}]{Stappers:2011}
{Stappers}, B.~W., et~al., Jun. 2011. {Observing pulsars and fast transients
  with LOFAR}. Astron. Astrophys. 530, A80.

\bibitem[{{Stewart} et~al.(in prep.)}]{Stewart:2014}
{Stewart}, A., et~al., in prep. {LOFAR NCP Transient Search}.

\bibitem[{{Sutherland} and {Saunders}(1992)}]{Sutherland:1992}
{Sutherland}, W., {Saunders}, W., Dec. 1992. {On the likelihood ratio for
  source identification}. Mon. Not. R. Astron. Soc. 259, 4130.

\bibitem[{{Swinbank}(2011)}]{Swinbank:2011}
{Swinbank}, J., Jul. 2011. {The LOFAR Transients Pipeline}. In: {Evans}, I.~N.,
  {Accomazzi}, A., {Mink}, D.~J., {Rots}, A.~H. (Eds.), Astronomical Data
  Analysis Software and Systems XX. Vol. 442 of Astronomical Society of the
  Pacific Conference Series. p. 313.

\bibitem[{{Swinbank}(2014)}]{Swinbank:2014}
{Swinbank}, J.~D., 2014. {Comet: A VOEvent broker}. Astron. Comp. 7, 12--26.

\bibitem[{{Szalay} et~al.(2006)}]{Szalay:2005}
{Szalay}, A., et~al., 2006. {Indexing the sphere with the Hierarchical
  Triangular Mesh}. MSR TR 2005 123, {The John Hopkins University, Microsoft
  Research and CERN}.

\bibitem[{{Taylor} et~al.(2012)}]{Taylor:2012}
{Taylor}, G.~B., et~al., Dec. 2012. {First Light for the First Station of the
  Long Wavelength Array}. Journal of Astronomical Instrumentation 1, 50004.

\bibitem[{{Thyagarajan} et~al.(2011)}]{Thyagarajan:2011}
{Thyagarajan}, N., et~al., Nov. 2011. {Variable and Transient Radio Sources in
  the FIRST Survey}. Astrophys. J. 742, 49.

\bibitem[{{Tingay} et~al.(2013)}]{Tingay:2013}
{Tingay}, S.~J., et~al., Jan. 2013. {The Murchison Widefield Array: The Square
  Kilometre Array Precursor at Low Radio Frequencies}. Publ. Astron. Soc. Aus.
  30, 7.

\bibitem[{{Usov} and {Katz}(2000)}]{Usov:2000}
{Usov}, V.~V., {Katz}, J.~I., Dec. 2000. {Low frequency radio pulses from
  gamma-ray bursts?} Astron. Astrophys. 364, 655--659.

\bibitem[{{van der Laan}(1966)}]{vanderLaan:1966}
{van der Laan}, H., Sep. 1966. {A Model for Variable Extragalactic Radio
  Sources}. Nature 211, 1131.

\bibitem[{{van der Tol} et~al.(2007){van der Tol}, {Jeffs}, and {van der
  Veen}}]{vdTol:2007}
{van der Tol}, S., {Jeffs}, B.~D., {van der Veen}, A.-J., Sep. 2007.
  {Self-Calibration for the LOFAR Radio Astronomical Array}. IEEE Transactions
  on Signal Processing 55, 4497--4510.

\bibitem[{{van Haarlem} et~al.(2013)}]{vanHaarlem:2013}
{van Haarlem}, M.~P., et~al., Aug. 2013. {LOFAR: The LOw-Frequency ARray}.
  Astron. Astrophys. 556, A2.

\bibitem[{{Welch} et~al.(2009)}]{Welch2009}
{Welch}, J., et~al., Aug. 2009. {The Allen Telescope Array: The first
  widefield, panchromatic, snapshot radio camera for radio astronomy and SETI}.
  IEEE Proceedings 97, 1438--1447.

\bibitem[{Whitaker(1996)}]{Whitaker:1996}
Whitaker, J.~C., 1996. The Electronics Handbook. CRC Press.

\bibitem[{{Williams} et~al.(2013)}]{Williams:2013}
{Williams}, P.~K.~G., et~al., Jan. 2013. {ASGARD: A Large Survey for Slow
  Galactic Radio Transients. I. Overview and First Results}. Astrophys. J. 762,
  85.

\bibitem[{{Williams} et~al.(2012){Williams}, {Barthelmy}, {Denny}, {Graham},
  and {Swinbank}}]{Williams:2012}
{Williams}, R.~D., {Barthelmy}, S.~D., {Denny}, R.~B., {Graham}, M.~J.,
  {Swinbank}, J., Sep. 2012. {Responding to the Event Deluge}. Vol. 8448 of
  SPIE Conf. Ser.

\bibitem[{Wolter(1984)}]{Wolter:1984}
Wolter, K.~M., 1984. An investigation of some estimators of variance for
  systematic sampling. Journal of the American Statistical Association
  79~(388), 781--790.
\newline\urlprefix\url{http://www.jstor.org/stable/2288708}

\end{thebibliography}

\end{document}